\documentclass[nojss,shortnames]{jss}

\usepackage{geometry} 
\usepackage{graphicx}
\usepackage{amsmath,scalerel}
\usepackage{amssymb}
\usepackage{setspace}
\usepackage{multirow}
\usepackage{array}
\usepackage{float}
\usepackage{caption}
\usepackage{subcaption}
\usepackage[normalem]{ulem}
\newcolumntype{P}[1]{>{\centering\arraybackslash}p{#1}}

\DeclareMathOperator*{\Bigcdot}{\scalerel*{\cdot}{\bigodot}}

\makeatletter 
\def\@eqnnum{{\normalsize \normalcolor (\theequation)}} 
\makeatother

\author{Waleed Almutiry\\Qassim University \And
Vineetha Warriyar K V\\ Uinversity of Calgary \And 
        Rob Deardon\\University of Calgary }

\title{Continuous Time Individual-Level Models of Infectious Disease: \pkg{EpiILMCT}}

\Plainauthor{Waleed Almutiry, Vineetha Warriyar K V, Rob Deardon}
\Plaintitle{Continuous Time Individual-Level Models of Infectious Disease: a Package EpiILMCT}
\Shorttitle{Continuous Time ILMs of Infectious Disease: \pkg{EpiILMCT}}

\Abstract{
This paper describes the \proglang{R} package \pkg{EpiILMCT}, which allows users to study the spread of infectious disease using continuous time individual level models (ILMs). 
The package provides tools for simulation from continuous time ILMs that are based on either spatial demographic, contact network, or a combination of both of them, and for the graphical summarization of epidemics. Model fitting is carried out within a Bayesian Markov Chain Monte Carlo (MCMC) framework. 
The continuous time ILMs can be implemented within either susceptible-infected-removed ($\mathcal{SIR}$) or susceptible-infected-notified-removed ($\mathcal{SINR}$) compartmental frameworks. As infectious disease data is often partially observed, data uncertainties in the form of missing infection times - and in some situations missing removal times - are accounted for using data augmentation techniques. 
The package is illustrated using both simulated and an experimental data set on the  spread of the tomato spotted wilt virus (TSWV) disease.
 }
\Keywords{\pkg{EpiILMCT}, infectious disease, individual level modelling, spatial models, contact networks, \proglang{R}}
\Plainkeywords{EpiILMCT, infectious disease, individual level modelling, spatial, contact network, R}

\Address{
  Waleed Almutiry\\
  Department of Mathematics\\
  College of Science and Arts in Ar Rass\\
  Qassim University\\
  Qassim, Saudi Arabia\\
  E-mail: \email{wkmtierie@qu.edu.sa}\\

  Vineetha Warriyar K V\\
  Sport Injury Prevention Research Centre\\
  Faculty of Kinesiology\\
  University of Calgary\\
  Calgary, AB Canada\\
  E-mail: \email{vineethawarriyar.kod@ucalgary.ca}\\
  
  Rob Deardon\\
  Faculty of Veterinary Medicine\\
  Department of Mathematics and Statistics\\
  University of Calgary\\
  Calgary, AB Canada\\
  E-mail: \email{robert.deardon@ucalgary.ca}\\
  URL: \url{http://people.ucalgary.ca/~robert.deardon/}\\
}

\begin{document}

\section[Introduction]{Introduction}

Innovative mathematical and mechanistic approaches to the modelling of infectious diseases are continuing to emerge in the literature. These can be used to understand the spread of disease through a population - whether homogeneous or heterogeneous - and enable researchers to construct predictive models to develop control strategies to disrupt disease transmission. For example, \citet{deardon2010inference} introduced a class of discrete time individual-level models (ILMs) which incorporate population heterogeneities by modelling the transmission of disease given various individual-level risk factors. The general framework of ILMs have already been successfully applied to a broad range of epidemic data, eg., the 2001 UK foot-and-mouth outbreak \citep{deardon2010inference,deeth2016spatial,malik2016parameterizing}, tomato spotted wilt virus (TSWV) disease \citep{pokharel2014supervised,pokharel2016gaussian}, the spread of 1-18-4 genotype of the porcine reproductive and respiratory syndrome in Ontario swine herds \citep{kwong2013bayesian}, and influenza transmission within households in Hong Kong during 2008 to 2009 and 2009 to 2010 \citep{malik2014individual}. Equivalent continuous time ILMs which capture the complex interactions between susceptible and infected individuals through spatial and contact networks can also be considered. The inference and fitting of such models is generally considered within a Bayesian framework using Markov chain Monte Carlo (MCMC). 

However, infectious disease epidemiologists have previously found it difficult to apply these individual-level models to real life problems. This is due to a dearth of readily available software products. The applicability of the aforesaid continuous time ILMs is implemented in an \proglang{R} \citep{CRAN} package, \pkg{EpiILMCT} \citep{EpiILMCT} and is available from Comprehensive \proglang{R} Archive Network (CRAN) at \url{https://CRAN.R-project.org/package=EpiILMCT}. In this article, we describe the package, \pkg{EpiILMCT} which allows users to simulate and fit epidemic data using distance- and/or network-based models \citep{bifolchi2013spatial,deardon2010inference,jewell2009bayesian}, and can also incorporate risk factors associated with both susceptible and infectious individuals. \pkg{EpiILMCT} also uses data augmentation techniques to carry out inference when the infection and/or removal times are unknown or censored, as is usually the case. To the extent of our knowledge, this feature is not available in any existing \proglang{R} packages that permit epidemic data analysis and modelling. Tools for the graphical summarization of epidemic data sets and outcomes are also provided. The statistical inferences made in \pkg{EpiILMCT} are set in a Bayesian framework and are carried out using Markov Chain Monte Carlo (MCMC). The main aim here is to provide a fast implementation of continuous time ILMs under different epidemic modelling frameworks. Because of the computationally intensive nature of MCMC for such models, we have coded functions, including MCMC, in \proglang{Fortran} to speed up computation. 

There are several \proglang{R} packages that permit a range of different modelling tools that allow for fitting spatial-temporal epidemic data. 
For example, the packages \pkg{splancs} \citep{splancs}, and \pkg{lgcp} \citep{lgcp1,lgcp2} provides methods for analyzing epidemic data as spatial and space-time point patterns. Also, the package \pkg{surveillance} \citep{surveillance2} implements a spatio-temporal point process model for epidemic data through the function \code{twinstim}. Other packages fit a range of autocorrelation regression spatio-temporal models (e.g., \pkg{CARBayesST} \citep{JSSCARBayesST}, \pkg{spdep} \citep{bivandcomparing1,bivandcomparing2}, and \pkg{spTimer} \citep{spTimer,JSSspTimer}). Further packages are mentioned in the Handling and Analyzing Spatio-Temporal Data CRAN task view \citep{Pebesma}. The \citet{RECON} provides further useful resources for disease outbreak analysis related \proglang{R} software packages. 

However, in each case, the functionality (e.g., models available) of the packages above is quite different to that of \pkg{EpiILMCT}. The models of the \pkg{EpiILMCT} package are ``mechanistic'' in that they attempt to more directly model the mechanisms of transmission between individuals.
Specifically, they take into account the spatial interactions between individuals with differing disease status (e.g., susceptible, infected, notified, removed) at continuous time points of the epidemic process. Those spatial interactions between susceptible and infectious individuals are incorporated as distance-based effects on the infectivity rate of individuals through an infection kernel function (power-law or Cauchy). The infectivity rates can also depend upon various susceptibility and transmissibility covariates at the individual level. 
Additionally, and of key importance, none of the aforementioned packages account for uncertainty in the event times using Bayesian data augmentation MCMC method.

There are several \proglang{R} packages that provide for the visualization, simulation and modelling the spread of epidemics through networks. 
The package \pkg{EpiModel} \citep{EpiModel} allows epidemic simulation from mathematical models of infectious disease through stochastic contact networks based on exponential-family random graph models (ERGMs). 
Some packages assume observed contact network or networks when fitting the specified model; for example, \pkg{ergm} \citep{ergm1,ergm2}, \pkg{Bergm} \citep{Bergm1}, and \pkg{hergm} \citep{hergm}. Those packages implement Bayesian analyses for fitting exponential-family transmission network models to observed contact network data. 

A recently developed package, \pkg{epinet} \citep{epinet2}, allows users to infer transmission networks from time-series epidemic data by modelling the contact network using a generalization of the ERGMs.
This package make use of time-series epidemic data as the input assuming unknown contact network in their functionality, and producing parameter estimates of the epidemic model as well as the contact and transmission networks. The transmission model can contain various covariates that captures important features (summary statistics) of the contact network as well as epidemic transmission.

However, once again these packages have different approaches to that implemented in \pkg{EpiILMCT}. We focus here on incorporating a contact network as a covariate in the implemented ILMs in \pkg{EpiILMCT}. The response in the ILMs is the event (e.g., infection) time, rather than the transmission network (the transmission network can be inferred later via posterior  predictive simulation, of course, but we do not address this here).
This is different to \pkg{epinet}, for example, which models the transmission network directly. 
The \pkg{EpiILMCT} package allows for any pre-user specified contact networks, including various special cases such as spatial or random unweighted (binary) (un)directed contact networks or weighted contact network.  

As both spatio-temporal and contact network-based mechanisms can be key to understanding the dynamics of infectious disease spread, the ILMs in \pkg{EpiILMCT} allow for the incorporation of both contact network and distance-based effects jointly in the infectivity rate of individuals. None of the aforementioned packages have this feature in their functionalities. 

The use of individual level data in more mechanistic epidemic models has been implemented in only a few other \proglang{R} packages. 
The most established of these is \pkg{surveillance} \citep{surveillance1,surveillance2}, a package for temporal and spatio-temporal disease modelling. It provides tools for outbreak detection in routinely collected surveillance data, as well as a range of models for infectious disease data. The most closely related model in \pkg{surveillance} to those of \pkg{EpiILMCT} is the additive endemic-epidemic multivariate temporal point process model.
These models are implemented in the \code{twinSIR} function for modelling the susceptible-infectious-recovered ($\mathcal{SIR}$) event history of a fixed population in continuous time using individual level data. 
However, not only is the underlying model framework different to that considered in the \pkg{EpiILMCT} package, but the \code{twinSIR} function does not allow for uncertainty in event times to be taken into account via data augmentation techniques. The function does not allow for only the epidemic terms of the model to be considered, as can be done in \pkg{EpiILMCT}; both endemic (e.g., seasonal) and epidemic terms must be included in the analysis. In addition, the distance kernel used in the epidemic part of the \code{twinSIR} function is represented by a linear combination of non-negative basis functions and is thus different from the distance kernels used in the \pkg{EpiILMCT} package. 

The \pkg{EpiILM} package \citep{EpiILM} that has recently been made available in \proglang{R}, provides similar utility to \pkg{EpiILMCT}, but for discrete-time ILMs. The models it contains provide options to include susceptible individual covariate information, as well as a choice to describe population heterogeneity. However, the package is limited to discrete-time distance-based or network-based infection kernels and requires known event histories (i.e., there is no data augmentation feature). 

As stated previously, inference for the models of \pkg{EpiILMCT} is carried out in a Bayesian MCMC framework.
Although there are packages available in \proglang{R} to implement MCMC algorithms such as \pkg{MCMCpack} \citep{MCMCpack} and \pkg{adaptMCMC} \citep{adaptmcmc}, all are based on random walk Metropolis-Hastings (M-H) algorithm. The data augmented MCMC algorithm used in the \pkg{EpiILMCT} package to fit various models uses random walk and independence sampler (within Gibbs) steps within a M-H algorithm. The independence sampler algorithm in our package appears to be essential for updating the missing data efficiently (event times and infectious periods), and the authors having not found it possible to achieve well-mixing MCMC chains if purely random walk M-H algorithms are used (even if tuned adaptedly). 

Our main purpose of developing this package is to make the use of continuous time ILMs available to epidemiologists and statisticians, through \proglang{R}, one of the most commonly used statistical software packages. Overall, \pkg{EpiILMCT} offers greatly increased flexibility for analyzing complex disease data.
The remainder of this paper is laid out as follows. In the next section, we describe the general continuous individual-level model implemented in \pkg{EpiILMCT}. We also discuss the different infection kernel functions implemented in the package. Sections 3 and 4 discuss the functions contained within the package and the underlying Bayesian inference, respectively. Section 5 illustrates the application of \pkg{EpiILMCT} to simulated and real data, while Section 6 concludes the paper with a short summary of the software package and its implications. 

\section{Model}\label{model}

The \pkg{EpiILMCT} package allows for the implementation of continuous time equivalents, and extensions, of the discrete-time individual-level models (ILMs) of \citet{deardon2010inference}. The compartmental frameworks considered are the susceptible-infectious-removed ($\mathcal{SIR}$) and susceptible-infectious-notified-removed ($\mathcal{SINR}$). In both frameworks, each individual is assumed to be in one of these states at any point in time, $t \in \mathbb{R}^{+}$. In the $\mathcal{SIR}$ framework, infected individuals transition between states, susceptible to infectious and from infectious to removed. 
Individuals are assumed to be in the susceptible ($\mathcal{S}$) state until they become infected at which point they become immediately infectious ($\mathcal{I}$), then being able to transmit the disease for the duration of their infectious periods before entering the removed ($\mathcal{R}$) state. In the $\mathcal{SINR}$ framework, infectious individuals are assumed to move from the infectious state ($\mathcal{I}$) to a notified ($\mathcal{N}$) state. The latter represents a state in which individuals have been identified as having the disease, and may be subjected to various restrictions (e.g., government-imposed movement constraints in the 2001 UK Foot-and-Mouth disease (FMD) outbreak). The $\mathcal{N}$-state infectivity rate is often assumed to be lower than that of $\mathcal{I}$-state. As infectious individuals enter the $\mathcal{R}$-state, they are removed from the infectious population (e.g., because of recovery and acquired immunity, death or quarantine) and from thereon play no role in transmitting the disease. 

%!!!!!!!!!!!!!!!!!!!!!!!!!!!!!!!!!!!!!!!!!!!!!!!

A full epidemic history consists of all transition event times for all individuals, and defines the state of all individuals at each point in time. For example for the $\mathcal{SINR}$ framework, $\mathcal{S}$(t), $\mathcal{I}$(t), $\mathcal{N}$(t) and $\mathcal{R}$(t) at time t for $t \in [0,t_{obs}]$ is defined by all infection, notification and removal times. Here, $t_{obs}$ is the maximum removal time; i.e., the time that the last notified individual enters the removed state. We assume that each susceptible individual $j$ at time t has an infectivity rate\footnote{Note that, technically the infectivity rates are conditioned upon the past epidemic history, so might be written $\lambda_{ij}(t|H_{t})$ where $H_{t}$ is the epidemic history up to time $t$. However, for the sake of brevity and simplicity we have dropped the conditioning from the notation.} with a given infectious individual $i$: 
\begin{equation}
\lambda_{ij}(t) = \left\{
  \begin{array}{l l}
    \lambda_{ij}^{-}(t)  & \quad i \in \mathcal{I}(t), j \in \mathcal{S}(t)\\
    \lambda_{ij}^{+}(t) & \quad i \in \mathcal{N}(t), j \in \mathcal{S}(t) \\
  \end{array} \right.,
\end{equation}
\noindent where
\[
\lambda_{ij}^{-}(t)  = \Omega_{S}(j) \Omega_{T}(i) \kappa(i,j)
\]
\[
\lambda_{ij}^{+}(t)  = \gamma \Omega_{S}(j) \Omega_{T}(i) \kappa(i,j), \quad \gamma > 0,
\]
\noindent where $\Omega_{S}(j)$ and $\Omega_{T}(i)$ are the susceptibility and transmissibility functions, respectively. They are defined as: 
\[ \Omega_{S}(j) = \mathbf{S} \mathbf{X}_{.j}^{\phi} \ \ \textrm{and}  \ \ \  \Omega_{T}(i) = \mathbf{T} \mathbf{Z}_{.i}^{\xi},  \quad  \phi,\xi >0, \]

\noindent where $\mathbf{S}$ and $\mathbf{T}$ are the (coefficient) parameter vectors of the susceptibility and transmissibility covariates with sizes equal to the number of susceptibility ($p_S$) and transmissibility ($p_T$) covariates, respectively; $\mathbf{X}^{\phi}_{.j}$ and $\mathbf{Z}^{\xi}_{.i}$ are the $j^{th}$ and $i^{th}$ columns of the susceptibility and transmissibility risk factor matrices $\mathbf{X}^{\phi} \in \Bbb R_{p_{S}\times n}^{+}$ and $\mathbf{Z}^{\xi} \in \Bbb R_{p_{T}\times n}^{+}$, respectively; and $\phi$ and $\xi$ are vectors of the power parameters of the susceptibility and transmissibility functions with sizes equal to $p_S$ and $p_T$, respectively. Note that, $\mathbf{X}^{\phi}$ and $\mathbf{Z}^{\xi}$ are constrained to be positive.
These power parameters allow for non-linearity between the susceptibility and transmissibility risk factors and the infection rate \citep{deardon2010inference}. The notification effect parameter $\gamma$ is used to measure the risk of infection after notification that can be reduced or increased depending on the disease type. For example, the transmissibility has been observed to increase after symptoms in SARS \citep{pitzer2007estimating}, whereas, it can be lower for the 2001 UK FMD \citep{jewell2009bayesian}. The latter stated this effect parameter in their general model as a control measure parameter that accounts only the reduction in the risk of infection. In the case of $\gamma=1$, notification has no effect on infectivity.

So, the total rate of infectivity of each susceptible individual $j$ at time $t$ is given by:
\begin{equation}
\lambda_{j}(t) = \left[\sum_{i \in \mathcal{N}^{-}{(t)}}{\lambda_{ij}^{-}(t)} + \sum_{i \in \mathcal{N}^{+}{(t)}}{\lambda_{ij}^{+}(t)} \right] + \epsilon(j,t),
\label{ratesinr}
\end{equation}
\noindent where $\mathcal{N}^{-}{(t)}$ is the set of infectious individuals at time $t$ who have been infected but have not reached the notified state; and $\mathcal{N}^{+}{(t)}$ is the corresponding set for notified individuals \citep{jewell2009bayesian}.

The nomenclature is the same for the $\mathcal{SIR}$ framework, but without the $\mathcal{N}$(t) state, there is not need to compartmentalize infectious individuals into pre- and post-notification sets. Therefore, the total rate of infectivity of each susceptible individual $j$ at time $t$ is given by:
\begin{equation}
\lambda_{j}(t) = \left[ \sum_{i \in \mathcal{I}{(t)}}{\lambda_{ij}^{-}(t)} \right]+ \epsilon(j,t) ,
\label{ratesir}
\end{equation}
\noindent where $\mathcal{I}{(t)}$ is the set of infectious individuals at time $t$ (i.e., they have been infected, but not yet removed).

The infectivity rate $\lambda_{j}(t)$ also contains a spark function that is denoted by $\epsilon(j,t)$ which allows for random infections otherwise unexplained by the model. This might represent, for example, the infection of a susceptible individual from a source outside of the observed population. In this model, we fix the spark term $\epsilon(j,t)$ such that $\epsilon(j,t) = \epsilon$; $\epsilon \geq 0$.

The infection kernel $\kappa(i,j)$ represents shared risk factors between pairs of infected and susceptible individuals. 
In the \pkg{EpiILMCT} package we consider three kernel types: distance-based, network-based, and combined distance and network-based. Two sub-types of distance-based kernel are also considered: Cauchy and power-law.
The infection kernel functions are given in Table~\ref{tab.kernel}. In the distance-based ILMs, the kernel function is based on the distances $d_{ij}$ between individuals generally, but not always, spatial Euclidean distance. In the network-based ILMs, the kernel function is based on the connections between individuals in a contact network that are represented by binary connections $c_{ij} =$ 0 or 1, or weighted connections $w_{ij}\in [0,\infty)$. In the combined ILMs the kernel consists of a linear function of both.

\begin{table}[t]
\centering
{\small
\begin{tabular}{ P{3.5cm}  |  P{3.8cm} | P{6cm} }
\hline 
&&\\
 Model  &  Kernel type  &  Kernel function \\ [2ex]
\hline \hline 
&&\\

  \multirow{2}{*}{Distance-based ILMs} 				&   Power-law  	& $\kappa(i,j) = d_{ij}^{-\beta}$, \hspace{0.3cm} $\beta>0$ \\ [2ex]
							   	&  Cauchy     	& $\kappa(i,j) = \frac{\beta}{d_{ij}^{2}+\beta^{2}}$,\hspace{0.3cm} $\beta>0$\\ [2ex]
\hline  
&&\\
  \multirow{2}{*}{Network-based ILMs} 				& Unweighted, undirected		& $\kappa(i,j) = c_{ij}$ ,\hspace{0.3cm} $c_{ij}=$ 0 or 1 	\\ [2ex]
                                     				& Weighted		& $\kappa(i,j) = w_{ij}$,\hspace{0.3cm} $w_{ij} \in[0,\infty)$ \\ [2ex]
\hline  
&&\\
	& \multirow{2}{*}{Power-law} 	& $\kappa(i,j) = d_{ij}^{-\beta_{1}} +\beta_{2}c_{ij}$  \\ [2ex]
Combined distance and network-based ILMs   	&  	& $\kappa(i,j) = d_{ij}^{-\beta_{1}} +\beta_{2}w_{ij}$ \\ [2ex]  \cline{2-3}
 &&\\	
					   			& \multirow{2}{*}{Cauchy}     	& $\kappa(i,j) = \frac{\beta_{1}}{(d_{ij}^{2}+\beta_{1}^{2})} + \beta_{2} c_{ij}$ \\ [2ex]
					   			&      	& $\kappa(i,j) = \frac{\beta_{1}}{(d_{ij}^{2}+\beta_{1}^{2})} + \beta_{2} w_{ij}$,\hspace{0.3cm} $\beta_{1},\beta_{2}>0$ \\ [2ex]
\hline
\end{tabular}
}
\caption{Types of kernel functions that are applied in the \pkg{EpiILMCT} package for fitting continuous time ILMs.}
\label{tab.kernel}
\end{table}

\subsection{Likelihood function}

We label the $m$ infected individuals $i = 1, 2, \dots, m$ with corresponding infection ($I_{i}$) and removal ($R_{i}$) times such that $I_{1} \leq I_{2} \leq \dots \leq I_{m}$. The $N-m$ individuals who remain uninfected after $t_{obs}$ are labeled $i = m + 1, m + 2, \dots, N$ with $I_{i}= R_{i} = \infty$. We then denote infection and removal time vectors for the population as ${\bf I} = \{I_{1}, \dots, I_{m}\}$ and ${\bf R} = \{R_{1}, \dots, R_{m}\}$, respectively. We assume that infectious periods follow a gamma distribution with a fixed shape $\delta_{a}$ and rate $\delta_{b}$, $\delta = (\delta_{a}, \delta_{b})$ \citep{jewell2009bayesian}. 
The likelihood function can be divided into two independent components: the infectious and the removed components. As we assumed earlier that each susceptible individual $j$ has a total infectivity rate $\lambda_{j}(I_{j})$ (their total specific infectious pressure) at the time of being infected ($I_{j}$) from infectious individuals $i \in \mathcal{I}(I_{j})$, the infectious component under the $\mathcal{SIR}$ continuous time ILMs can be written as:

{\small
\begin{eqnarray}
L_{1}&=&\prod_{j=2}^{m}{\left(\epsilon +\sum_{i:I_{i} < I_{j} \leq R_{i}}{\lambda_{ij}^{-}(I_{j})} \right)}\nonumber\\
&\times& \exp \left\{ -\int_{I_{1}}^{t_{obs}}{\left(\sum_{i \in \mathcal{S}(u)}{\epsilon}  +  \sum_{i \in \mathcal{I}(u)}\sum_{j \in \mathcal{S}(u)}{\lambda_{ij}^{-}(u-I_{i})}  \right) du} \right\}\nonumber 
\end{eqnarray}
}

\noindent where the product term represents the total specific infectious pressure that each infected individual receives from infectious individuals at the time of being infected, and the exponential integral represents the total person-to-person infectious pressure during the course of the epidemic. 

The removed component then contains the contribution of the infectious periods to the likelihood function via their densities. As the infectious period of an infected individual $i$ ($\mathcal{D}_{i} = R_{i} - I_{i}$) is independent of others, the removed component is simply:

{\small
\begin{eqnarray}
L_{2} &=& 
\prod_{i=1}^{m}{f(\mathcal{D}_{i};\delta)} \nonumber
\end{eqnarray}
}

The likelihood function of the general $\mathcal{SIR}$ continuous time ILMs can then be formed by combining the infectious and removal parts given as follows:

{\small
\begin{eqnarray}
L(\boldsymbol{I},\boldsymbol{R}|\boldsymbol{\theta}) &=& L_{1} \times L_{2} \nonumber\\
&=& 
\prod_{j=2}^{m}{\left(\epsilon +\sum_{i:I_{i} < I_{j} \leq R_{i}}{\lambda_{ij}^{-}(I_{j})}\right)}   \exp \left\{ -\sum_{i=1}^{m}{\left(\sum_{j=1}^{N}{( (R_{i} \wedge I_{j}) - (I_{i} \wedge I_{j})) \lambda_{ij}^{-}(I_{j})}\right)} \right\}  \nonumber \\
&\times& \exp \left(- \epsilon \sum_{i=1}^{N}{\left[(t_{obs} \wedge I_{i}) - I_{1}\right]} \right) \prod_{i=1}^{m}{f(\mathcal{D}_{i};\delta)} \hspace{3cm} \delta > 0,
	 \label{eq:eqsir}
\end{eqnarray}
}

\noindent where the wedge symbol $\wedge$ denotes the minimum operator; $\boldsymbol{\theta}$ is the vector of unknown parameters; $f(\Bigcdot;\delta)$ indicates the density of the infectious period distribution; and $\mathcal{D}_{i}$ is the infectious period of infected individual $i$ defined as $\mathcal{D}_{i}= R_{i}-I_{i}$. The integral in Equation~\ref{eq:eqsir}, which represents the total person-to-person infectious pressure through the course of the epidemic, can be written as the double sum in the lower equation \citep{britton2002bayesian,jewell2009bayesian}. The integral is transformed by discretizing it into a sum over the successive events of the epidemic and is substituted by the double sum. The likelihood function of the general $\mathcal{SINR}$ continuous time ILMs can be formed in a very similar manner (see Appendix~\ref{appendix.2}). 

\section[Contents of the EpiILMCT package]{Contents of the \pkg{EpiILMCT} package}

The \pkg{EpiILMCT} package can be used to simulate and graphically summarize epidemics, and, for a given model, carry out Bayesian inference and calculate log-likelihood. Most of the main package functions are written in \proglang{Fortran 95} (called from within the \proglang{R} wrapper), since they are computationally intensive tasks. The functions contained in the package are reviewed in Table~\ref{content}. 

\begin{table}[t]
\centering
\begin{tabular}{l|p{9.5cm}}
 \hline
 Function & Usage \\ [1ex]
 \hline\hline
\code{contactnet}		& Generates undirected unweighted (binary) contact network matrices from spatial
(\code{powerlaw}, or \code{Cauchy}), or \code{random}, network models. \\ [1ex]
\code{plot.contactnet}	& Provides plot of a contact network of class \code{`contactnet'}. \\ [1ex]
\code{datagen}			& Generates epidemics from distance/network-based individual level models.\\ [1ex]
\code{as.epidat}		& Generates objects of class \code{`datagen'} that contain the
individual event history of an epidemic along with other individual level information. \\ [1ex]
\code{plot.datagen}		& Provides different plots summarizing an epidemic of class \code{`datagen'}. \\ [1ex]
\code{epictmcmc}		& Runs a Bayesian data augmented MCMC algorithm for fitting specified models ($\mathcal{SIR}$ or $\mathcal{SINR}$).\\ [1ex]
\code{print.epictmcmc}		& Prints the contents of \code{`epictmcmc'} object to the console.\\ [1ex]
\code{summary.epictmcmc}		& Summary method for \code{`epictmcmc'} objects.\\ [1ex]
\code{plot.epictmcmc}		& Plots the output of \code{`epictmcmc'} object.\\ [1ex]
\code{loglikelihoodepiILM} & Calculates the log likelihood for a given compartmental framework and kernel type of the continuous time ILMs. \\ [1ex]
\hline
\end{tabular}
\caption{Description of functions and their usages in the \pkg{EpiILMCT} package.}
\label{content}
\end{table}%

\subsection{Contact network}

Various types of contact network can be considered. First, we consider unweighted (binary) contact networks which can be directed or undirected. In an undirected unweighted contact network, each pair of individuals share the same symmetric connection such that $c_{ij}=c_{ji}$ for $i \ne j$; $i$, $j = 1, \dots, N$; and each network is defined by $\binom{N}{2}$ elements where $c_{ij}$ = 1 if a connection exists between individuals $i$ and $j$, and 0 otherwise. In a directed unweighted contact network, it is not necessary for individuals to share the same symmetrical relationship so that $c_{ij} \ne c_{ji}$ for $i \ne j$; $i$, $j = 1, \dots, N$. This leads to a non-symmetric contact network matrix. Weighted contact networks can also be considered in the \pkg{EpiILMCT} package in which the connections between individuals are not described as present or absent but are weighted according to their strength. These too can be directed or undirected. 

A function (\code{contactnet}) is included to generate undirected unweighted contact networks. It can simulate both spatial networks where connections are more likely to occur between individuals closer in space (``spatial contact networks''), as well as random contact networks. 
The function \code{contactnet} has three available options (\code{"powerlaw"}, \code{"Cauchy"}, and \code{"random"}) for the network model, where the first two options simulate spatial contact networks in which the probability of connections between individuals are based on required XY coordinate input. 

The inclusion of the two options \code{"powerlaw"} and \code{"Cauchy"} in the argument \code{type} is to allow the user to choose between two commonly assumed spatial forms to describe the underlying population.
For example, the power-law network model is taken from \citet{bifolchi2013spatial} who use this network to test how well purely spatial power-law ILMs can approximate disease spread through networks.  The Cauchy model was used by \citet{jewell2009bayesian} to model the 2001 UK foot-and-mouth outbreak in Cumbria; they found this kernel the most appropriate for predicting transmission of these tested. 

We now describe the three model options in detail. First, in the power-law contact network model of \citet{bifolchi2013spatial} the probability of a connection between individual $i$ and $j$ is given by: 
\[ p(c_{ij}=1) = 1- e^{-\nu(d_{ij}^{-\beta})}, \quad  \nu,\beta > 0, \]
\noindent where $d_{ij}$ is the Euclidean distance between individuals $i$ and $j$;  $\beta$ is the spatial parameter; and $\nu$ is the scale parameter. 

Under the Cauchy contact network model, as used in \citet{jewell2009bayesian}, the probability of a connection between individual $i$ and $j$ is given by:
\[ p(c_{ij}=1) = 1- e^{-\beta/(d_{ij}^{2} + \beta^{2})} , \quad \beta > 0, \]
\noindent where $d_{ij}$ is the Euclidean distance between individuals $i$ and $j$; and $\beta$ is the spatial parameter.

Finally, under the random contact network model, the probability of a connection is simply generated from a Bernoulli distribution with probability equal to $\beta$.

Let us now consider some examples. To create the above undirected unweighted contact networks, the function requires the network model to be specified (\code{"powerlaw"}, \code{"Cauchy"}, or \code{"random"}) via the \code{type} argument. If \code{"powerlaw"} or \code{"Cauchy"} are selected, the XY coordinates of individuals (\code{location}) have to be specified through the argument \code{location}. 
The function \code{contactnet} produces a list which includes the contact network matrix in a class, \code{`contactnet'}.

To obtain a plot of the contact network, we introduce an \proglang{S3} method \code{plot.contactnet} function, which uses as its input an object of the class \code{`contactnet'}. The \code{plot.contactnet} function uses code internal to \pkg{EpiILMCT} for the layout when plotting power-law or Cauchy network models, but depends on the package \pkg{igraph} \citep{igraph} when plotting random network model.

The following code generates the three types of contact networks for a population of 50 individuals, with a uniformly distributed spatial layout for the spatial network models.

\begin{Sinput} 
R> library("EpiILMCT")
R> set.seed(12345)
R> loc <- matrix(cbind(runif(50, 0, 10), runif(50, 0, 10)), ncol = 2,
+    nrow = 50)
R> net1 <- contactnet(type = "powerlaw", location = loc, beta = 1.5,
+    nu = 0.5)
R> net2 <- contactnet(type = "Cauchy",  location = loc, beta = 0.5)
R> net3 <- contactnet(type = "random", num.id = 50, beta = 0.08)
R> par(mfrow=c(2, 2))
R> plot(net1)
R> plot(net2)
R> plot(net3, xlab = "(random)", vertex.color = "red", vertex.size = 20,
+    edge.color = "black", vertex.label.cex = 0.5, 
+    vertex.label.color = "black")
\end{Sinput}

\begin{figure}[!h]
\begin{center}
\includegraphics[width=13cm,height=9cm]{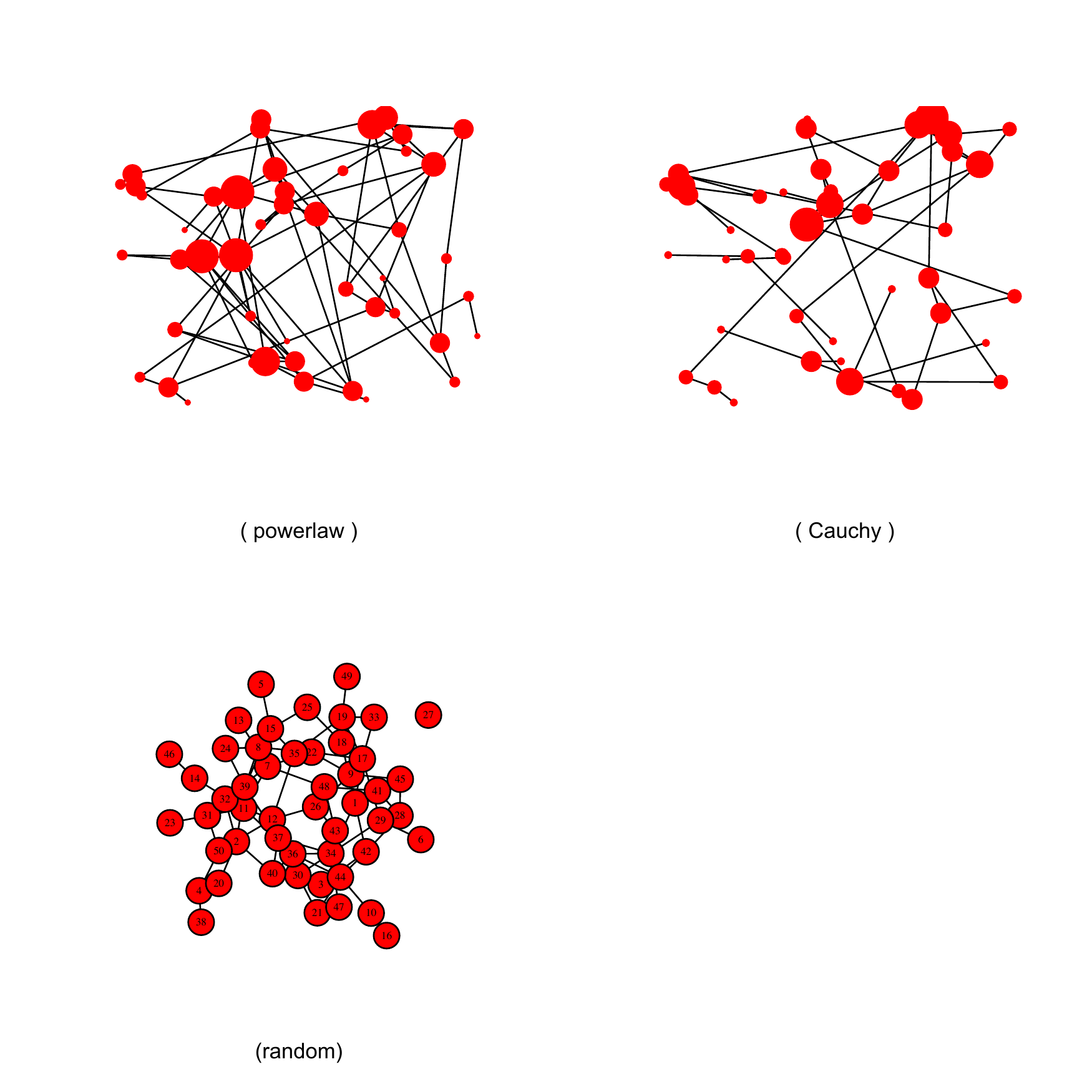}
\caption{Examples of the three undirected unweighted (binary) contact network models generated for the same population. Red dots represent nodes with size corresponding to their degree (number of edges). }
\label{fig.network-plots}
\end{center}
\end{figure}

A realization of the three networks for a given population is shown in Figure~\ref{fig.network-plots}. Note the underlying spatial layout of the nodes is the same for both spatial network models.

\subsection{Epidemic simulation}

The function \code{datagen} allows the user to generate epidemics from the continuous time ILMs under the $\mathcal{SIR}$ or $\mathcal{SINR}$ compartmental frameworks. Which framework is to be used is specified through the \code{type} argument. Each infected individual in a simulated epidemic has an infection life history defined by their time of infection and the length of time spent in the infectious state. We assume the conditional intensity functions stay constant between events, such that the time to the next infection, given that the last infection occurred at time $t$, follows $W_{j} \sim$ Exp($\lambda_{j}(t)$). Here, $W_{j}$ represents the ``waiting time'' for susceptible individual $j$ becoming infected. 

Under the $\mathcal{SIR}$ framework, and using the chosen distribution of the infectious period, an epidemic is simulated starting with a randomly chosen initial infected individual $k$ at time $I_{1} = 0$, or with initial infected individual(s) specified via the argument \code{initialepi}. This argument requires a vector or matrix containing the id number(s), removal time(s), infectious period(s) and infection time(s) of the infected individual(s). At time $I_{s}$, the waiting time until infection for susceptible individual $j$ is then drawn from $W_{j} \sim$ Exp($\lambda_{j}(I_{s})$).

The individual with the minimum $W$ is taken as the next infected individual and assigned an infection time $I_{s+1} = I_{s}+min(W)$; an infection period $\mathcal{D}_{j}$ (generated from $f(\mathcal{D}_{j};\delta))$; and a removal time $R_{s+1} = I_{s+1}+\mathcal{D}_{j}$. The process is repeated until no infectives remain in the population or $I_{s+1} > t_{max}$, where $t_{max}$ is the time at where the epidemic simulation is set to end. $t_{max}$ can be then specified via the option \code{tmax}.

Under the $\mathcal{SINR}$ framework, each infected individual is considered to have an incubation period comprising the time from infection to notification, and a delay period comprising the time from notification to removal. Together the incubation and delay periods constitute the infectious period. An epidemic is simulated in the same manner described above for the $\mathcal{SIR}$ framework, except that the infection period is replaced by incubation and delay periods $\mathcal{D}^{(inc)}$ and $\mathcal{D}^{(delay)}$ (generated from $f(\mathcal{D}^{(inc)}_{j};\delta^{(inc)})$ and $f(\mathcal{D}^{(delay)}_{j};\delta^{(delay)})$, respectively); and notification and removal times are assigned as $N_{s+1} = I_{s+1}+\mathcal{D}^{(inc)}_{j}$ and $R_{s+1} = N_{s+1}+\mathcal{D}^{(delay)}_{j}$, respectively.

In this function, the infectious, incubation and delay periods are assumed to follow either exponential or gamma distributions. These distributions can be specified through the \code{delta} argument. Under the $\mathcal{SIR}$ framework, \code{delta} is a vector containing the shape and rate parameters of a gamma distribution, whereas under the $\mathcal{SINR}$ framework it is a 2$\times$2 matrix where each row represents the parameters of the incubation and delay period distributions. Note that - as is often done - an exponential distribution can be assigned to any of these distributions by setting shape parameter equal to one.

The epidemic data structure output of the \code{datagen} function is used throughout the \pkg{EpiILMCT} package. Under an $\mathcal{SIR}$ ILM, it returns a matrix with four columns representing: the id numbers of the individuals, removal times, infectious periods, and infection times. Under an $\mathcal{SINR}$ ILM, it returns a matrix with six columns: the id numbers of the individuals, removal times, delay periods, notification times, incubation periods, and infection times. Uninfected individuals are assigned infinity values ($Inf$) for both their removal and infection times. Epidemic data from other modelling packages can be extracted and modified to be used in \pkg{EpiILMCT}. For example, we show how this can be done using the individual level models from the \pkg{surveillance} package in Appendix~\ref{appendix.1}.

The choice of kernel function $\kappa(i,j)$ is specified using the \code{kerneltype} argument. This takes one of three options: \code{"distance"} for distance-based, \code{"network"} for network-based, or \code{"both"} for distance and network-based. The appropriate kernel matrix must also be provided via the \code{kernelmatrix} argument. If \code{"distance"} is chosen as the \code{kerneltype}, the user must choose a spatial kernel (\code{"powerlaw"} or \code{"Cauchy"}) through the \code{distancekernel} argument. The distance matrix can be obtained from XY coordinate data using the \code{dist} function from the \pkg{stats} package \citep{CRAN}. Otherwise the distance matrix can be specified by the user. Other arguments in the \code{datagen} function require the data and coefficient parameters for the susceptibility and transmissibility risk factors as explained in Section~\ref{model}. 

We define an object of class \code{`datagen'} to take a list of values needed for the use of other functions, such as, \code{plot.datagen} and \code{epictmcmc}. This list contains: \code{type}, \code{kerneltype}, \code{epidat} (event times), \code{location} (XY coordinates of individuals), and \code{network} (contact network matrix). In the case of setting the \code{kerneltype} to \code{"distance"}, a NULL value will be assigned to the \code{network} option. The package has also a separate function \code{as.epidat} that generates an object of class \code{`datagen'} for a given epidemic data set (Appendix~\ref{appendix.1} contains a brief example of using this function).

The package also contains an \proglang{S3} method \code{plot.datagen} function, which illustrates disease spread through the epidemic timeline. This function can be used for either distance-based or network-based ILMs. The object of this function has to be of class \code{`datagen'}. If the \code{plottype} argument is set to \code{"history"}, the function produces epidemic curves of infection and removal times. Example plots are shown in Figure \ref{fig.network-history}. Conversely, setting this argument to \code{"propagation"} produces plots of the epidemic propagation over time. With the latter option, exactly which plots are output varies by kernel. With the network kernel, the function plots all the connections between individuals and overlays these with the epidemic pathway direction over time. This path direction consists of directed edges from all infectious individuals connected to a given newly infected individual $i$ with infection time $I_{i}$ (one per plot). Thus, this produces directed networks showing possible pathways of the disease propagation. 
With the distance kernel, the function plots the spatial epidemic dispersion over time. It shows the changes in the individual status that related to the chosen compartmental framework. To avoid displaying too many plots, the \code{time.index} argument allows user to obtain propagation plots at specific infection time points rather than at every infection time. 

\section{Bayesian inference}\label{Bayesian Inference}

\begin{figure}[!h]
\begin{center}
\includegraphics[width=14cm,height=9cm]{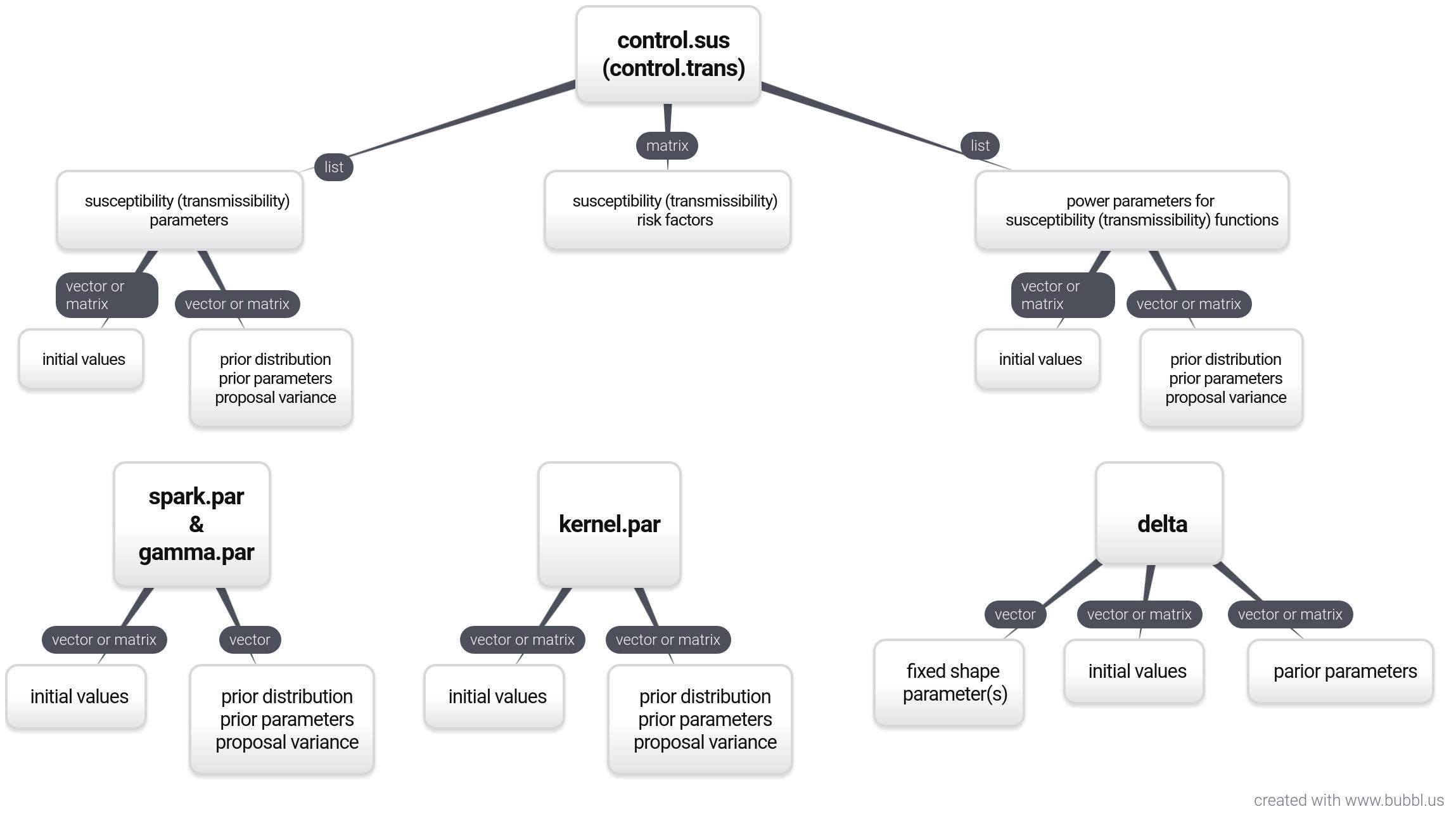}
\caption{A diagram of the input structure for the arguments \code{control.sus}, \code{control.trans}, \code{kernel.par}, \code{spark.par}, \code{gamma.par} and \code{delta} in the function \code{epictmcmc}.}
\label{diagram-epictmcmc}
\end{center}
\end{figure}

Prior distributions of the model parameters are selected from one of three options: gamma, positive half normal or uniform distribution. Then, Metropolis-Hastings MCMC is performed to estimate the joint posterior of the model parameters and latent variables (the latter if various event times are assumed unknown). This is achieved using the function \code{epictmcmc}. The parameters of the susceptibility and transmissibility functions, infection kernel and spark term (collectively denoted $\boldsymbol{\theta}$) are updated using the random-walk proposals. The user is required to tune the proposal variances to achieve good mixing properties. Thus, the user must provide a vector of initial values, a prior distribution (\code{"gamma"}, \code{"uniform"}, or \code{"halfnormal"}), the prior parameters, and the variance of the normal proposal distribution for each parameter as shown in Figure~\ref{diagram-epictmcmc}. In case of running multiple MCMC chains, the user should provide a vector of initial values of the model parameters. Note that, setting the variance of the normal proposal distribution to zero fixes a parameter at its initial value. This option allows the user to fix such a parameter in the model while updating others (i.e., conditioning on the parameters).

Using the \code{datatype} argument, the \code{epictmcmc} function allows for three scenarios in terms of event time uncertainty: \code{"known epidemic"} can be used to model a fully observed epidemic with known infection and removal times; \code{"known removal"} can be used to model a partially observed epidemic where the infection times are unknown; and \code{"unknown removal"} can be used to model a partially observed epidemic where removal and infection times are unknown. The latter option is only available for the $\mathcal{SINR}$ continuous time ILMs where notification times are assumed correctly known. When the \code{datatype} argument is set to \code{"known epidemic"}, the infectious periods are fixed by default.

When infection times are unknown, the rate(s) of the infectious, incubation and/or delay period distributions are assigned gamma prior distributions with shape a and rate b. Thus, the rate parameters have conditional distributions with a standard form following the gamma distribution. For the $\mathcal{SIR}$ continuous time ILMs, this is as follows: 
\[ \delta|\boldsymbol{\theta},\boldsymbol{I}, \boldsymbol{R} \sim \Gamma(m+a_{\delta},M+b_{\delta}), \]
\noindent where $\delta$ is the rate of the infectious period distribution; $M=\sum_{i=1}^{m}{(R_{i}-I_{i})}$; and $a_{\delta}$ and $b_{\delta}$ are the prior parameters of the infectious period rate. For the $\mathcal{SINR}$ continuous time ILMs, the distribution of the incubation rate and delay parameters are as follows: 
\[ \delta^{(inc)}|\boldsymbol{\theta},\boldsymbol{I},\boldsymbol{N}, \boldsymbol{R} \sim \Gamma(m+a_{\delta^{(inc)}},M_{inc}+b_{\delta^{(inc)}}), \]
\noindent where  $\delta^{(inc)}$ is the rate of the incubation period distribution; $M_{inc}=\sum_{i=1}^{m}{(N_{i}-I_{i})}$; and $a_{\delta^{(inc)}}$ and $b_{\delta^{(inc)}}$ are the prior parameters of incubation period rate; and
\[ \delta^{(delay)}|\boldsymbol{\theta},\boldsymbol{I},\boldsymbol{N}, \boldsymbol{R} \sim \Gamma(m+a_{\delta^{(delay)}},M_{delay}+b_{\delta^{(delay)}}), \]
\noindent where $\delta^{(delay)}$ is the rate of the delay period distribution; $M_{delay}=\sum_{i=1}^{m}{(R_{i}-N_{i})}$; and $a_{\delta^{(delay)}}$ and $b_{\delta^{(delay)}}$ are the prior parameters of delay period rate.

A Gibbs update (i.e., sampling from the conditional posterior distribution) is used for the infectious period rate (for the $\mathcal{SIR}$ continuous time ILMs) or the incubation and/or delay period rates (for the $\mathcal{SINR}$ continuous time ILMs). The required information for each period distribution are entered via the \code{delta} argument. We assume each period type follows a gamma distribution with fixed shape and unknown rate. Thus, to update the rate parameter of each period we specify \code{delta}, a list containing a vector of the fixed shape value(s), a vector (matrix) of the initial values of the rate(s), and a vector (matrix) for the parameters of the prior distribution of the rate parameter(s). In the case of incubation and delay periods being estimated, the input of the initial values is a $2 \times$ \code{nchains} matrix, and the prior parameters is a $2\times2$ matrix where each row contains the required information for each period rate.

An independence sampler is then used to update the infection times/infectious periods (for the $\mathcal{SIR}$ continuous time ILMs), or the infection times/incubation periods and/or the removal times/delay periods (for the $\mathcal{SINR}$ continuous time ILMs). For the $\mathcal{SIR}$ continuous time ILMs, the $i^{th}$ infection time $I_{i}$ is updated by generating an infectious period $\mathcal{D}^{*}_{i}$ from a gamma proposal distribution such that $\mathcal{D}^{*}_{i}\sim \Gamma(a,b)$. Then, the new infection time is the difference between the observed removal time and the new infectious period of the $i^{th}$ individual. The same procedure is used for updating the missing event times, infectious periods and corresponding parameters for the $\mathcal{SINR}$ continuous time ILMs. The parameter values of the gamma proposal distribution could be provided through the \code{periodproposal} argument. If they are not provided, the parameters of the gamma proposal distribution are then based on the fixed shape and updated rate values from the argument \code{delta}.
Computationally, it may be more efficient to apply a block update for the periods and event times. This can be implemented using the \code{blockupdate} argument, which requires that the user specifies $m$ (assuming removal and infection times are known for the first $m$ individuals), and the size of each block.

The \code{epictmcmc} function allows for sampling from multiple MCMC chains. This is done by providing the number of chains to be run via the option \code{nchains}. Additionally, multiple chains can be run in parallel by setting \code{parallel = TRUE}. This implies the use of the \code{parLapply} function from the \pkg{parallel} package \citep{CRAN}. The number of cores to be used is set to the minimum of the number of chains and the available cores on the user's computer. Note that, if \code{parallel} is set to FALSE and \code{nchains}>1, multiple MCMC chains are run sequentially. When \code{parallel} is set to TRUE, the \code{clusterSetRNGStream} function from the \pkg{parallel} package \citep{CRAN} is used to distribute the setting seed value by the \code{set.seed} function \citep{CRAN} to each core to reproduce the same results, otherwise each core sets its seed value from the current seed of the master process.

%For reproducing the same results, the random number generator of the \code{epictmcmc} \proglang{Fortran} function can be initialized to a fixed seed value through the option \code{seedval}. 

The output of this function is an object of class \code{`epictmcmc'}. 
There are \proglang{S3} methods: \code{print.epictmcmc}, \code{summary.epictmcmc} and \code{plot.epictmcmc} that depend on the \pkg{coda} package \citep{coda}.
The latter function has a \code{plottype} argument to specify which samples need to be plotted. This argument has three options: \code{"parameter"} to produce trace plots of the posterior distributions of the model parameters, and \code{"inf.times"} (\code{"rem.times"}) to produce plots of the average posterior and 95\% CI of the unobserved infection (removal) times when \code{datatype} set to \code{"known removal"} (\code{"unknown removal"}). The \proglang{S3} \code{plot.epictmcmc} method has the same options as the \code{plot.mcmc} function in the \pkg{coda} package, for example, \code{start}, \code{thin}, and \code{density}.

The class \code{`epictmcmc'} contains the MCMC samples of the model parameters and the missing information (if \code{datatype} is not set to \code{"known epidemic"}) as an \code{mcmc} matrix, and other useful information to be used in other functions, such as the above \proglang{S3} methods. So standard summary methods from \pkg{coda}, such as \code{summary.mcmc} and \code{plot.mcmc}, can be employed using these MCMC samples as inputs.

Posterior predictive checks of the fitted model can be performed using the \code{datagen} function. This requires that the user supplies the model parameter values with a combined sample of the MCMC model parameter outputs. If desired, the simulation can be constrained to the first $m$ infected individuals and their event times. This can be achieved by appending this information to the \code{initialepi} option. 

\section{Examples}

\subsection{Simulated network-based epidemic}

In this section, we illustrate the \pkg{EpiILMCT} package by fitting a simple $\mathcal{SIR}$ network-based continuous time ILM to a simulated epidemic. We consider an isolated population of 50 individuals distributed uniformly in an area of 10$\times$10 units. We also consider a binary susceptibility covariate $z$ which can be thought as being, say, an individual's treatment or vaccination status. Thus, the infectivity rate given in Equation~\ref{ratesir} becomes: 
\[
\lambda_{j}(t) = (\alpha_{0} + \alpha_{1}z_{j}) \sum_{i \in \mathcal{I}(t)}{c_{ij}}, \hspace{1cm} \alpha_{0}, \alpha_{1}>0,
\]
\noindent where the susceptibility function $\Omega_{S}(j) = \alpha_{0} + \alpha_{1}z_{j}$
; there are no transmissibility covariates $\Omega_{T}(i) = 1$; and $\epsilon = 0$. First, let us simulate the XY coordinates of individuals and the binary covariate $z$ as follows: 
\begin{Sinput}
R> set.seed(91938)
R> loc <- matrix(cbind(runif(50, 0, 10), runif(50, 0, 10)), ncol = 2,
+    nrow = 50)
R> cov <- cbind(rep(1, 50), rbinom(50, 1, 0.5))
\end{Sinput}

\begin{figure}[!h]
\begin{center}
\includegraphics[width=0.9\textwidth,height=9cm]{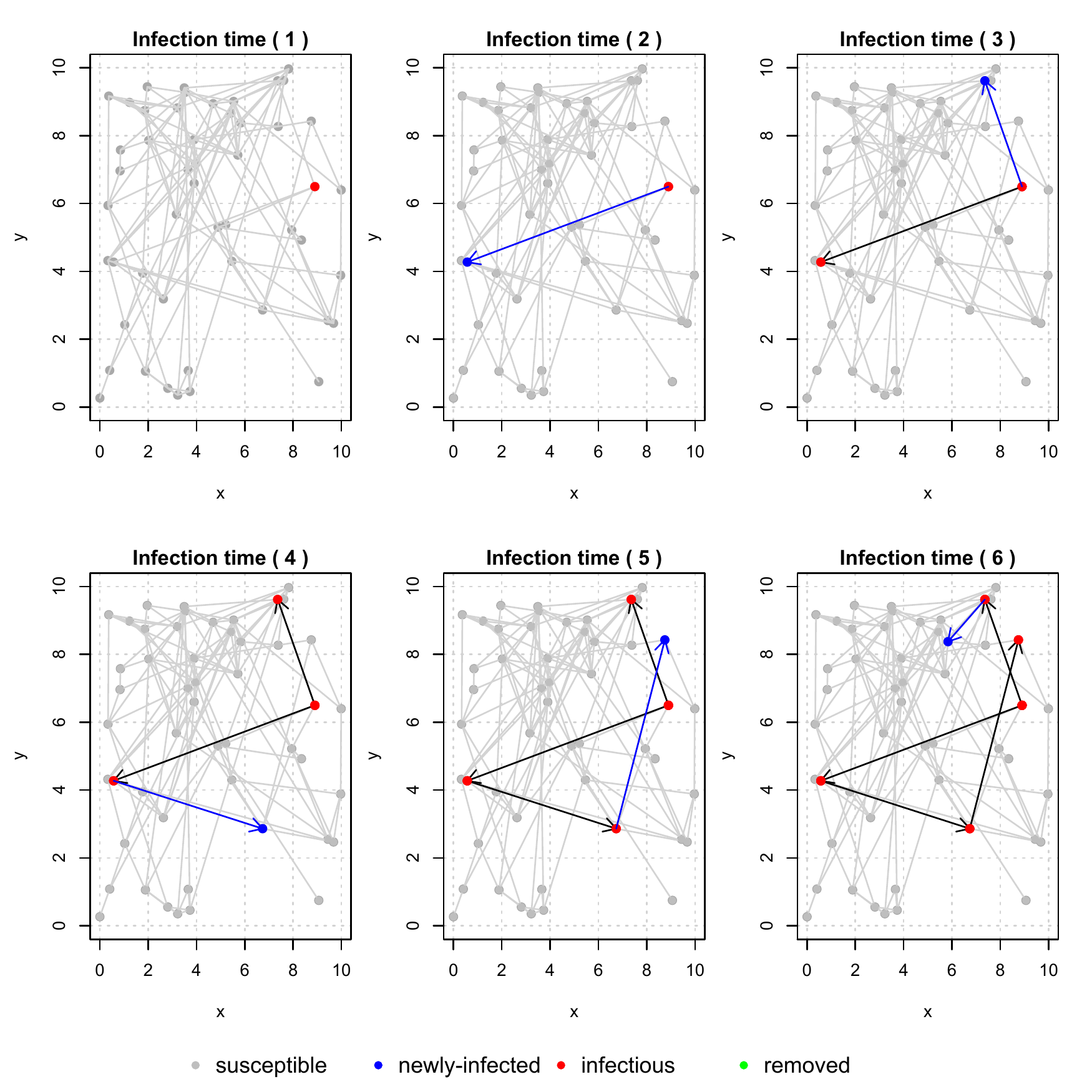}
\caption{The directed pathway network of the generated epidemic over its time-line using the simple network-based ILMs.}
\label{fig.network-trans}
\end{center}
\end{figure}

To simulate the epidemic, we generate a contact network using the \code{contactnet} function. Here, we use the power-law contact network model with $\beta = 1.8$ and $\nu = 1$, as illustrated in the following code: 
\begin{Sinput}
R> net <- contactnet(type = "powerlaw", location = loc, beta = 1.8,
+    nu = 1)
\end{Sinput}
Figure \ref{fig.network-trans} shows the contact network (grey lines).  The epidemic is then generated using the \code{datagen} function. Here, the epidemic is initialized with a randomly chosen infectious individual; then generated by providing the function with the contact network matrix, the susceptibility covariate and the following parameter values: $\alpha_{0} = 0.08$, $\alpha_{1} = 0.5$, and $\mathcal{D}_{i} \sim \Gamma(4,\delta = 2)$. This is coded as follows:

\begin{Sinput}
R> epi <- datagen(type = "SIR", kerneltype = "network", 
+    kernelmatrix = net, suspar = c(0.08, 0.5), delta = c(4, 2), 
+    suscov = cov)
\end{Sinput}

The object $epi$ is stored in the data file \code{NetworkData} as a class \code{`datagen'}, along with the susceptibility covariate ($cov$), available in the \pkg{EpiILMCT} package.

\begin{Sinput}
R> data("NetworkData", package = "EpiILMCT")
R> class(NetworkData[[1]])
\end{Sinput}
\begin{Soutput}
[1] "datagen"
\end{Soutput}
\begin{Sinput}
R> names(NetworkData[[1]])
\end{Sinput}
\begin{Soutput}
[1] "type"       "kerneltype" "epidat"     "location"   "network"   
\end{Soutput}
\begin{Sinput}
R> head(NetworkData[[1]]$epidat)
\end{Sinput}
\begin{Soutput}
     id.individual rem.time inf.period  inf.time
[1,]            50 1.526078  1.5260782 0.0000000
[2,]            16 2.612491  1.9933013 0.6191893
[3,]             5 2.394094  1.6567882 0.7373061
[4,]            45 3.169602  2.2370141 0.9325876
[5,]            44 1.805656  0.5661341 1.2395222
[6,]            19 1.737867  0.4576725 1.2801945
\end{Soutput}

To illustrate the propagation of the epidemic, we set the argument \code{plottype} to \code{"propagation"}. To limit the number of plots, we assign the \code{time.index} option to be a vector containing time points for plots to be generated as shown in the following code:

\begin{Sinput}
R> plot(NetworkData[[1]], plottype = "propagation", 
+    time.index = seq_len(6))
\end{Sinput}

We can also produce density plots of the infection and removal times, and a plot of the infectious periods, by specifying the argument \code{plottype} to \code{"history"} as shown in the following code:

\begin{Sinput}
R> plot(NetworkData[[1]], plottype = "history")
\end{Sinput}

\begin{figure}[!h]
\begin{center}
\includegraphics[width=0.9\textwidth,height=9cm]{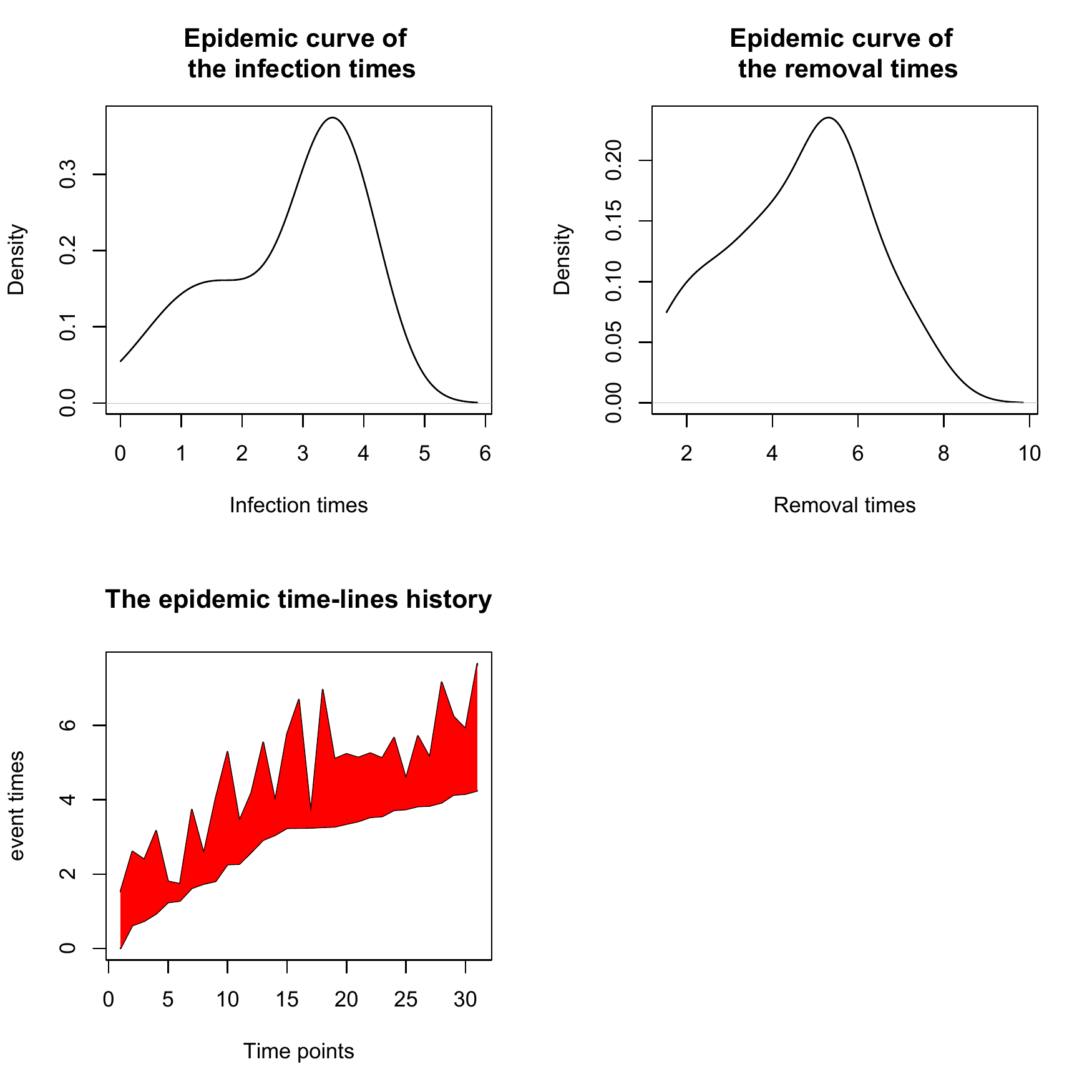}
\caption{The epidemic curves of the infection and removal times for the epidemic that was generated using the simple network-based continuous time ILM. The red shaded area in the third plot represent the infectious periods.}
\label{fig.network-history}
\end{center}
\end{figure}

Figure~\ref{fig.network-history} shows the densities of the infection and removal times, and the infectious periods; while Figure~\ref{fig.network-trans} shows the epidemic propagation plot.

To illustrate fitting continuous time ILMs to data, we analyze the epidemic using the \code{epictmcmc} function. This is done under two observation scenarios: \code{"known epidemic"} and \code{"known removal"}. For the former analysis, we assign $\Gamma(1,0.1)$ gamma prior distributions to the model parameters $\alpha_{0}$ and $\alpha_{1}$ and use normal MCMC proposals with variances equal to 0.5 and 1, respectively. 
As we have two susceptibility parameters, 
the argument \code{control.sus} is then a list that contains: 1) a list of a vector of initial values of $\alpha_{0}$ and $\alpha_{1}$, and a $2 \times 4$ matrix in which each row represents the required information for updating each parameter; and 2) a $50 \times 2$ matrix of the covariates representing the unity intercept and the binary covariate $z$.
Now, we run the MCMC using the \code{epictmcmc} function for sampling a single chain of 150,000 iterations using the following code: 

\begin{Sinput}
R> set.seed(91938)
R> suscov <- list(NULL)
R> suscov[[1]] <- list(c(0.01, 0.1), 
+    matrix(c("gamma", "gamma", 1, 1, 0.1, 0.1, 0.5, 1), ncol = 4, nrow = 2))
R> suscov[[2]] <- NetworkData[[2]]
R> mcmc1 <- epictmcmc(object = NetworkData[[1]], datatype = "known epidemic",
+    nsim = 150000, control.sus = suscov)
\end{Sinput}

The estimates of the model parameters can be obtained through the \proglang{S3} \code{summary} function of \code{epictmcmc}.  
The posterior means and 95\% credible intervals of these parameters can be obtained via the following command:
\begin{Sinput}
R> summary(mcmc1, start = 10000, thin = 10)
\end{Sinput}
\begin{Soutput}
********************************************************* 
Model: SIR network-based continuous-time ILM 
Method: Markov chain Monte Carlo (MCMC) 
Data assumption: fully observed epidemic 
number.chains : 1 chains 
number.iteration : 140000 iterations 
number.parameter : 2 parameters 

********************************************************* 
 1. Empirical mean and standard deviation for each variable,
plus standard error of the mean:
                Mean        SD    Naive SE Time-series SE
Alpha_s[1] 0.0850579 0.0268504 0.000226919    0.000298624
Alpha_s[2] 0.5082012 0.1290994 0.001091050    0.001179665
 2. Quantiles for each variable:
                2.5%       25%       50%      75%    97.5%
Alpha_s[1] 0.0417253 0.0655198 0.0824374 0.101868 0.143758
Alpha_s[2] 0.2856068 0.4163682 0.4982712 0.587971 0.789077
 3. Empirical mean, standard deviation, and quantiles for the log likelihood,
          Mean             SD       Naive SE Time-series SE 
   -55.8040938      1.0188095      0.0086102      0.0104071 

    2.5%      25%      50%      75%    97.5% 
-58.5949 -56.1864 -55.4943 -55.0810 -54.8118 
 4. acceptance.rate : 
Alpha_s[1] Alpha_s[2] 
  0.112361   0.222748 
\end{Soutput}

The MCMC trace plots for  the model parameters can be produced using the \proglang{S3} method \code{plot.epictmcmc}.
\begin{Sinput}
R> plot(mcmc1, plottype = "parameter", start = 10000, thin = 10, 
+    density = FALSE)
\end{Sinput}

\begin{figure}[!h]
\begin{center}
\includegraphics[width=0.9\textwidth,height=6cm]{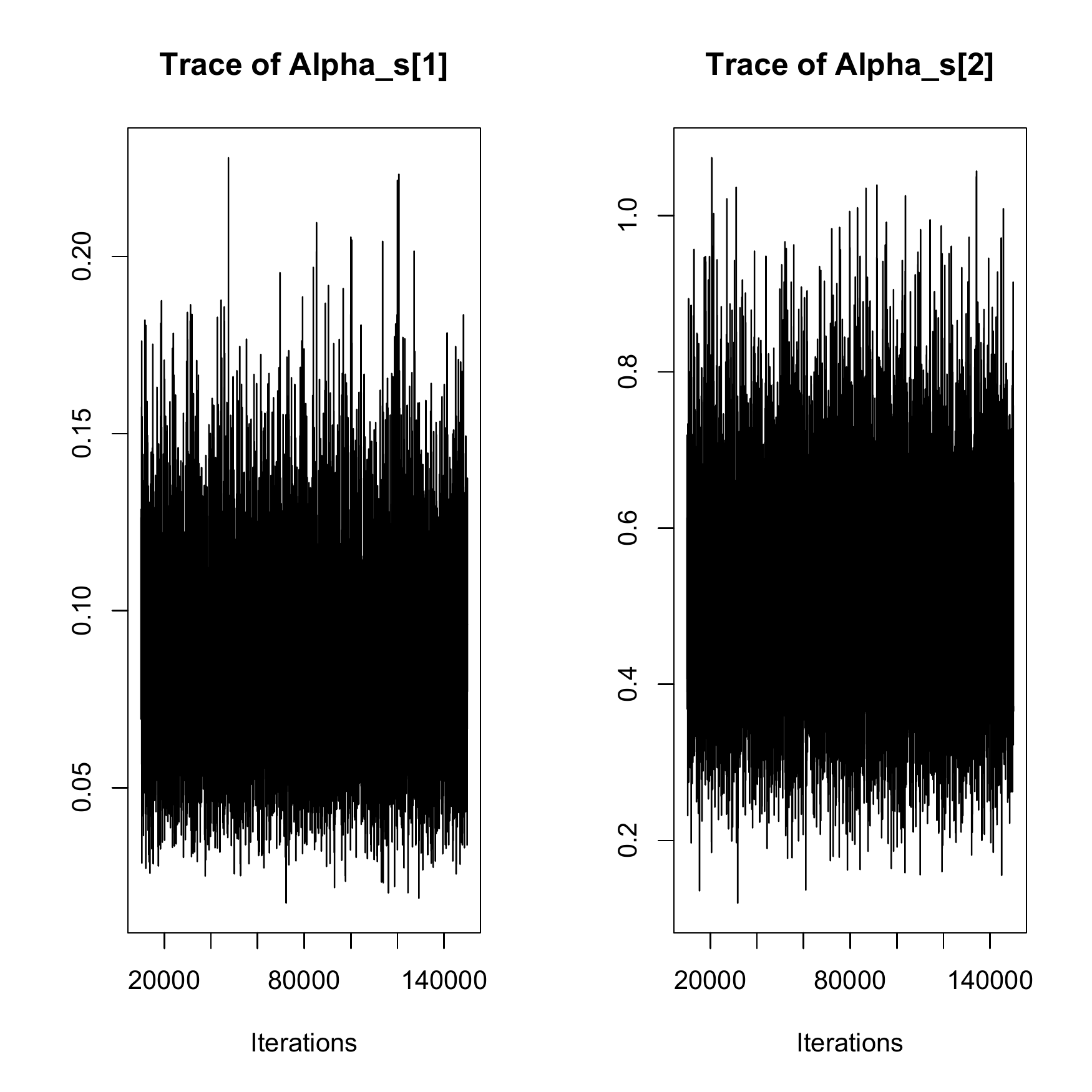}
\caption{The MCMC chains of the posterior distributions of the model parameters fitting the simulated epidemic using the network-based continuous time ILM assuming fully observed epidemic.}
\label{fig.network-mcmc-1}
\end{center}
\end{figure}

Figure~\ref{fig.network-mcmc-1} shows the MCMC trace plots for the model parameters $\alpha_{0}$ and $\alpha_{1}$. We observe a posterior mean of $\hat{\alpha}_{1}$ = 0.508 with 95\% credible (percentile) interval (0.286,0.789) and a posterior mean of $\hat{\alpha}_{0}$ = 0.085 with 95\% credible interval (0.042,0.144). We also observed well-mixed MCMC chains for both model parameters. 
The computation time for running the above MCMC code was 16 seconds on an Apple MacBook Pro with i5-core Intel 2.4 GHz processors with 8 GB of RAM.

For the known removal times analysis, \pkg{EpiILMCT} uses data augmented MCMC to infer infection times and the infectious period rate. Here, we assume that the infectious period follows a gamma distribution with shape 4 and unknown rate parameter $\delta$; so $\mathcal{D}_{i} \sim \Gamma(4, \delta)$. Here, we also include a spark term $\epsilon$. This is not strictly necessary but tends to improve MCMC mixing. We assigned gamma prior distribution $\Gamma(4,2)$ for $\delta$ and an exponential prior distribution with rate 0.01 for $\epsilon$.

We can then run the MCMC using the \code{epictmcmc} function for sampling a single chain of 150,000 iterations using the following code:

\begin{Sinput}
R> set.seed(91938)
R> suscov <- list(NULL)
R> suscov[[1]] <- list(c(0.01, 0.1), matrix(c("gamma", "gamma", 1, 1, 0.1, 
+    0.1, 0.2, 0.8), ncol = 4, nrow = 2))
R> suscov[[2]] <- NetworkData[[2]]
R> spark <- list(0.01, c("gamma", 1, 0.01, 0.1 ))
R> mcmc11 <- epictmcmc(object = NetworkData[[1]], datatype = "known removal", 
+    nsim = 150000, control.sus = suscov, spark.par = spark, delta = list(4, 
+    2, c(4, 2 )))
\end{Sinput}

\begin{figure}[!h]
\begin{center}
\includegraphics[width=0.9\textwidth,height=9cm]{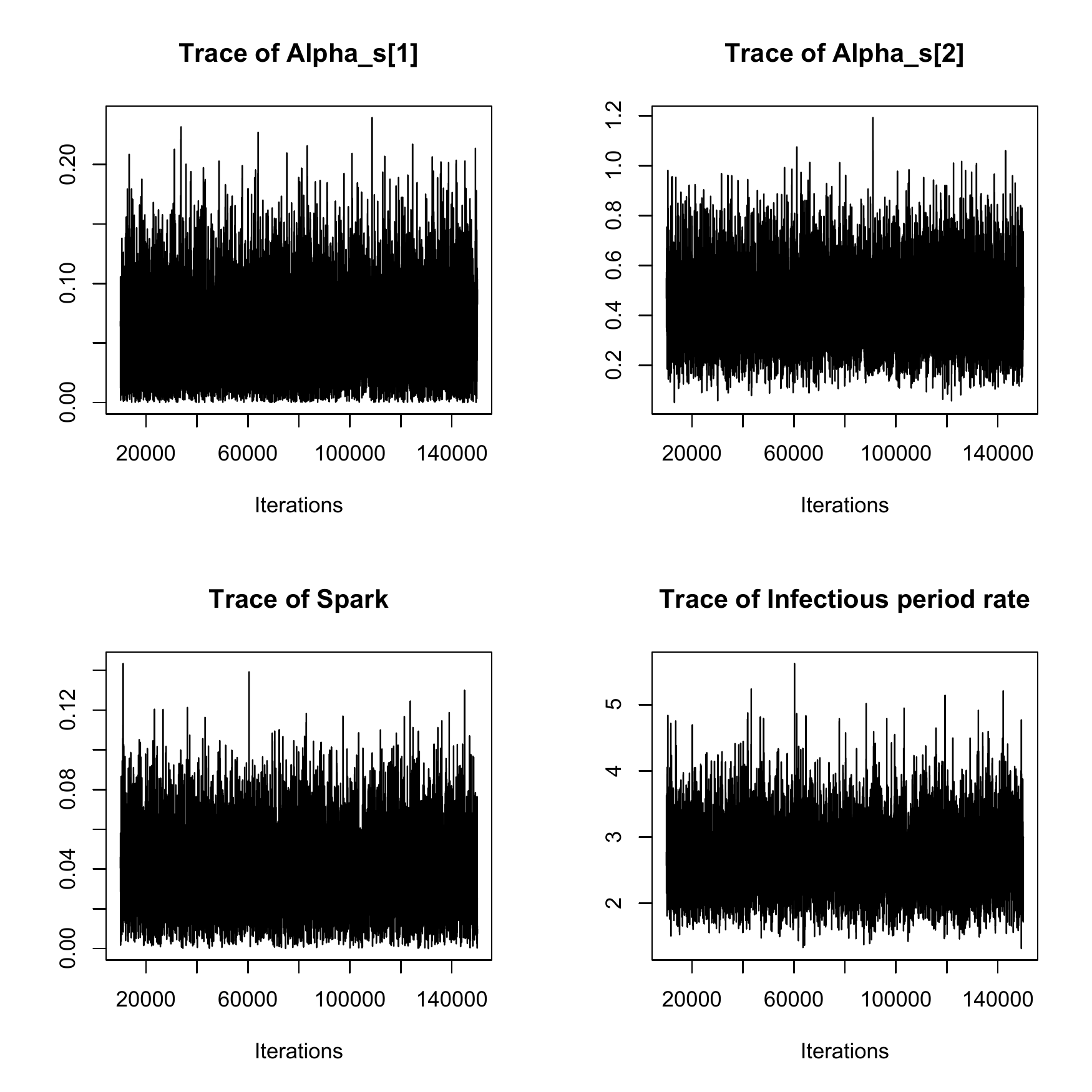}
\caption{The MCMC chains of the posterior distributions of the model parameters for fitting the simulated epidemic using the network-based continuous time ILM assuming partially observed epidemic (unknown infection times).}
\label{fig.network-mcmc-2}
\end{center}
\end{figure}

The computation time for the above code on the aforementioned machine was 201 seconds. Figure \ref{fig.network-mcmc-2} shows typical MCMC trace plots for the model parameters $\alpha_{0}$, $\alpha_{1}$, $\epsilon$, and $\delta$. Well-mixed MCMC chains are observed for all the model parameters. 

As the posterior samples of the model parameters are stored in the \code{epictmcmc} object as an \code{`mcmc'} object of the type used in the \pkg{coda} package, the standard \code{summary} methods from \pkg{coda} can be employed, inserting \code{mcmc1$parameter.samples} as the input of this function. This is illustrated in the following command:
\begin{Sinput}
R> summary(window(mcmc11$parameter.samples, start = 10000, thin = 10))
\end{Sinput}
\begin{Soutput}
Iterations = 10000:150000
Thinning interval = 10 
Number of chains = 1 
Sample size per chain = 14001 
1. Empirical mean and standard deviation for each variable,
   plus standard error of the mean:
                          Mean      SD  Naive SE Time-series SE
Alpha_s[1]             0.05717 0.03497 0.0002955      0.0004005
Alpha_s[2]             0.42976 0.14320 0.0012102      0.0015706
Spark                  0.03742 0.02001 0.0001691      0.0002553
Infectious period rate 2.64673 0.49444 0.0041786      0.0074440
2. Quantiles for each variable:
                           2.5%     25%     50%     75%   97.5%
Alpha_s[1]             0.004402 0.03059 0.05322 0.07817 0.13770
Alpha_s[2]             0.189793 0.32899 0.41698 0.51633 0.75099
Spark                  0.004977 0.02246 0.03548 0.04999 0.08208
Infectious period rate 1.821692 2.29301 2.59927 2.94168 3.73862
\end{Soutput}

\begin{figure}[!h]
\begin{center}
\includegraphics[width=0.9\textwidth,height=9cm]{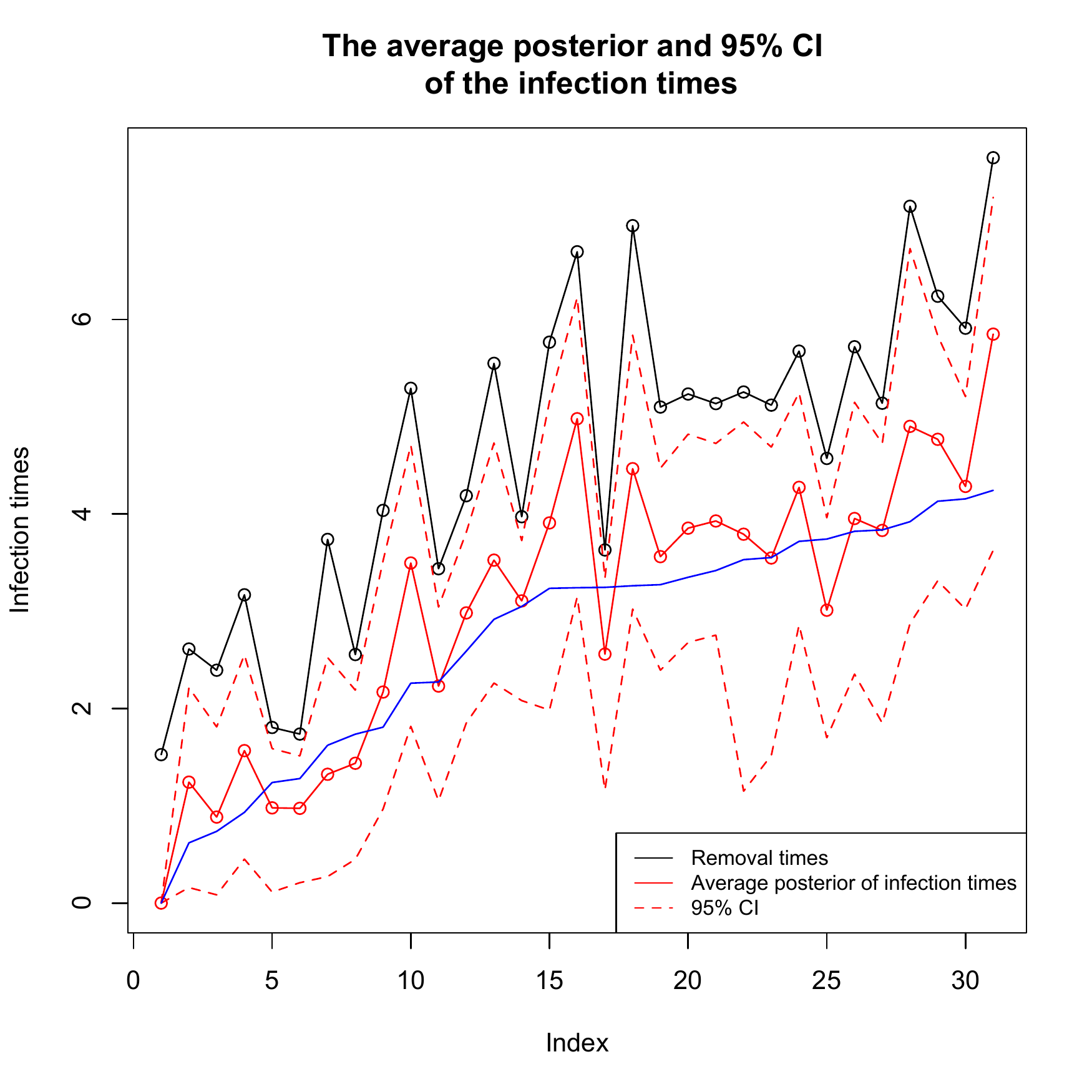}
\caption{The posterior means and 95$\%$ credible intervals of the infection times for fitting the simulated epidemic using the network-based continuous time ILM assuming partially observed epidemic (unknown infection times). The black line represent the observed removal times, solid red line represent the posterior means, dotted red lines represent the 95\% credible interval, and the blue line represents the observed infection times.}
\label{fig.network-mcmc-2-epi-curve}
\end{center}
\end{figure}

Thus, the posterior means and 95\% credible intervals of the model parameters are as follows: $\hat{\alpha}_{0} = 0.057$ (0.004,0.138), $\hat{\alpha}_{1} = 0.430$ (0.190,0.751), $\hat{\epsilon} = 0.037$ (0.005,0.082), and $\hat{\delta} = 2.647$ (1.822,3.739). The infection times are also well-approximated (see Figure~\ref{fig.network-mcmc-2-epi-curve}). Figures \ref{fig.network-mcmc-2} and \ref{fig.network-mcmc-2-epi-curve} are produced using the following code:
\begin{Sinput}
R> plot(mcmc11, plottype = "parameter", start = 10000, thin = 10, density = 
+    FALSE)
R> plot(mcmc11, epi = NetworkData[[1]], plottype = "inf.times", start = 
+    10000, thin = 10)
R> lines(NetworkData[[1]]$epidat[,4], type = "l", col = "blue")
\end{Sinput}

To check the fit of the model, we consider the posterior predictive distribution of four statistics. Specifically, we consider: $T_{1}$, the total number of infected individuals; $T_{2}$, the average removal time; $T_{3}$, the variance of the removal times; and $T_{4}$, the length of the epidemic. Here, we simulate 10,000 epidemics based on random draws of the model parameters from the MCMC output (excluding burnin) of the known removal times analysis (i.e., unknown infection times). We condition our simulation on the first ten infected individuals, then calculated the four statistics for each simulation. This simulation procedure is implemented in parallel using the \code{future_lapply} function from the \pkg{future.apply} package \citep{future.apply} as follows:

\begin{Sinput}
R> set.seed(524837)
R> mb <- sample(seq(10000, 150000), 10000)
R> posterior.pred <- function(x) {
+    epi <- datagen(type = "SIR", kerneltype = "network", kernelmatrix = 
+         NetworkData[[1]]$network, initialepi = 
+         matrix(NetworkData[[1]]$epidat[1:10, ], ncol = 4, nrow = 10), 
+         suspar = c(mcmc11$parameter.samples[x, 1], 
+         mcmc11$parameter.samples[x, 2]), spark = 
+         mcmc11$parameter.samples[x, 3], delta = c(4, 
+         mcmc11$parameter.samples[x, 4]), suscov = NetworkData[[2]])$epidat
+    numinf <- sum(epi[, 2] != Inf )
+    muremtime <- mean(epi[1:numinf, 2])
+    varremtime <- var(epi[1:numinf, 2])
+    lengthepi <- max(epi[1:numinf, 2])   
+    result <- c(numinf, muremtime, varremtime, lengthepi)
+    return(result)
+    }
R> library("future.apply")
R> plan(multiprocess, workers = 4) ## Parallelize using 
R> datmcmc <- future_lapply(mb, FUN = posterior.pred, future.seed = TRUE)
R> summary.results <- sapply(datmcmc, unlist, simplify = TRUE)
\end{Sinput}

\begin{figure}[!t]
\begin{center}
\includegraphics[width=0.9\textwidth,height=9cm]{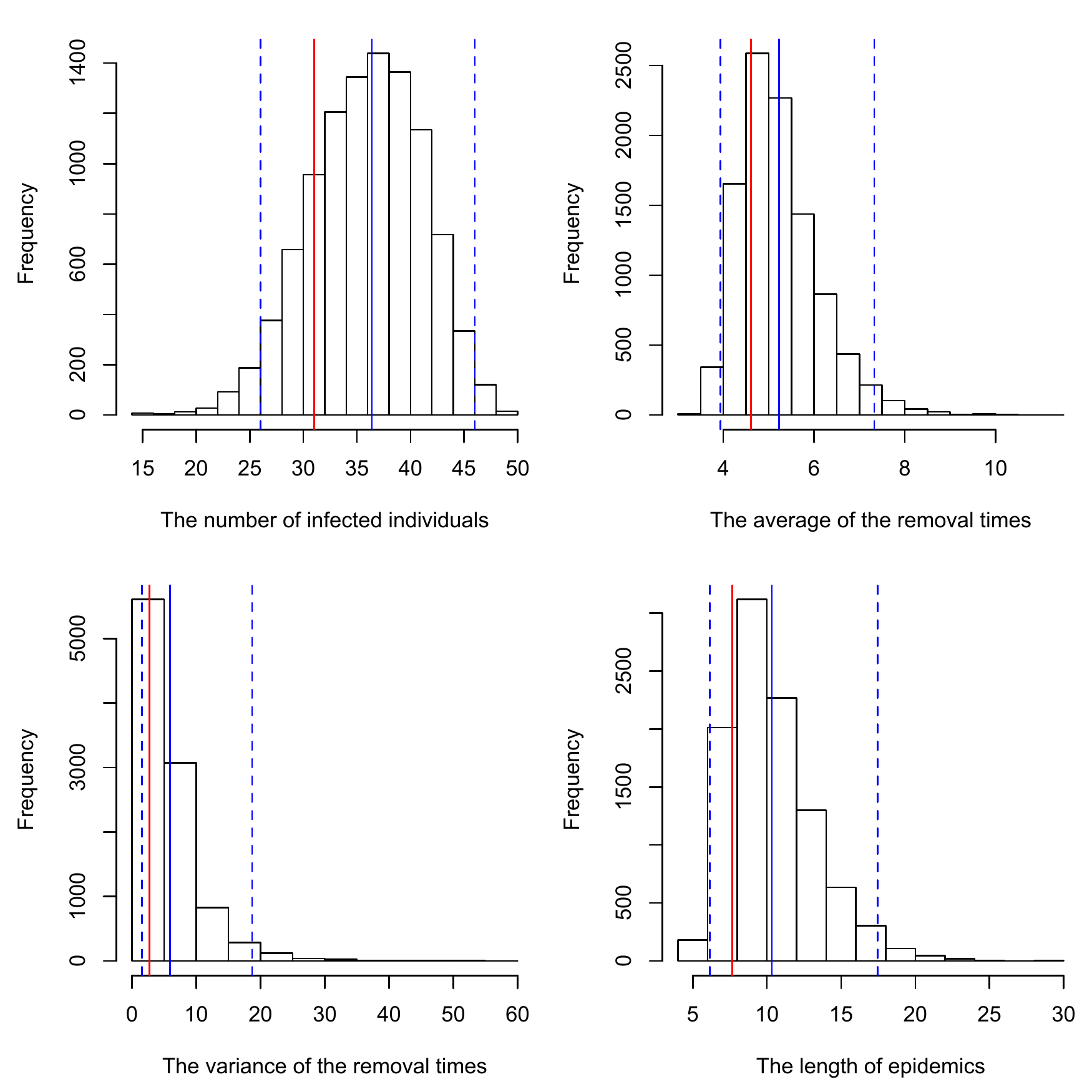}
\caption{The posterior predictive distributions of four statistics: The number of infected individuals, the average removal times, the variance of removal times, and the length of epidemic for fitting partially observed epidemic (unknown infection times) using network-based continuous time ILM. The red vertical lines represent the observed statistic values and the solid and dotted blue vertical lines represents the posterior predictive means and 95\% credible intervals of the four statistics.}
\label{fig.prediction unknown infection}
\end{center}
\end{figure}

The posterior predictive distributions of the four statistics are shown in Figure~\ref{fig.prediction unknown infection}. We can see that each distribution captures the observed statistics well.

\subsection{Case study: Tomato spotted wilt virus (TSWV) data}

\begin{figure}[!h]
\begin{center}
\includegraphics[width=10cm,height=7cm]{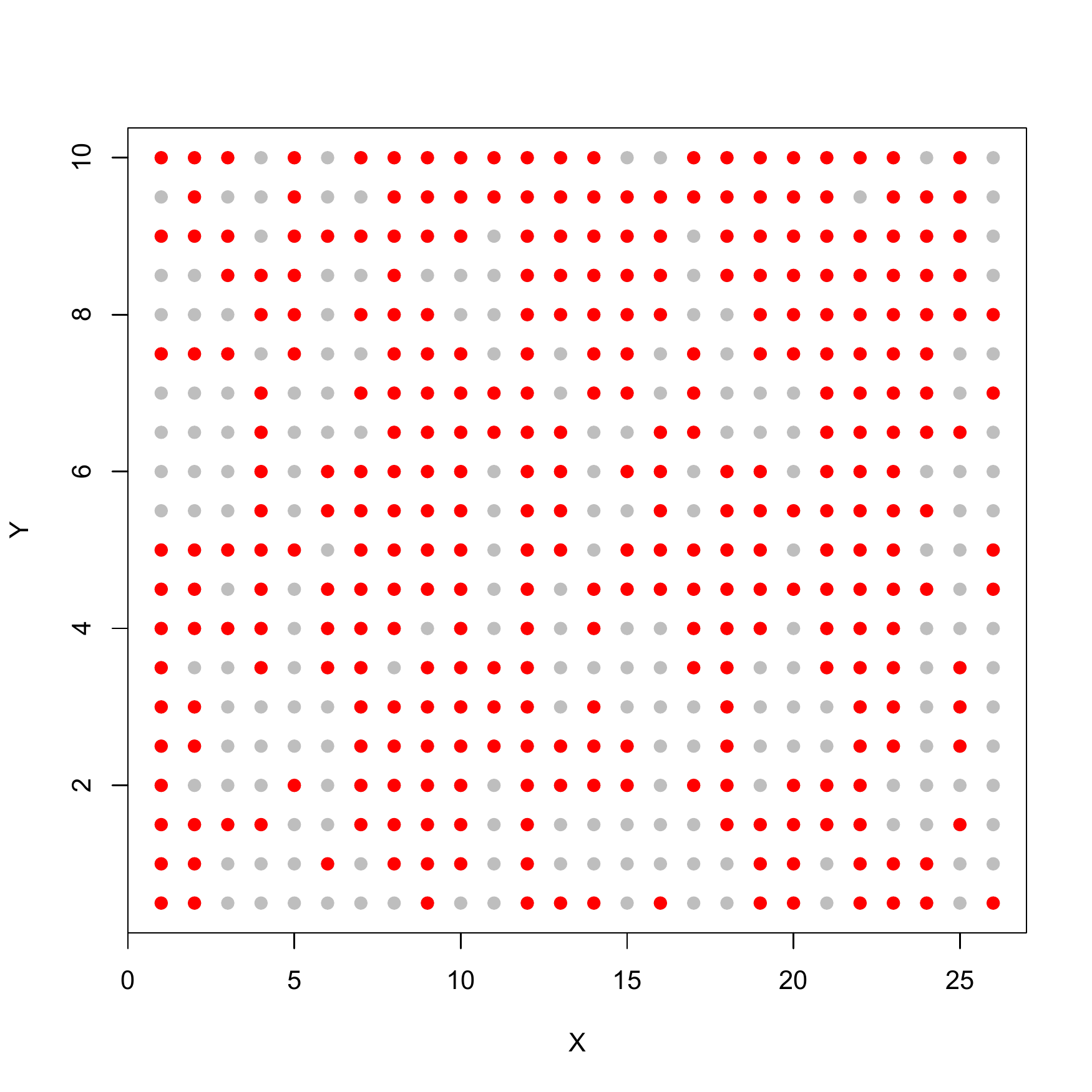}
\caption{A grid plot of the XY coordinates of plants of TSWV data. The red plots represent infected plants at the end of the disease.}
\label{TSWV}
\end{center}
\end{figure}

We further illustrate the \pkg{EpiILMCT} package by analyzing the TSWV data as described in \citet{hughes1997validating} and analyzed with spatial ILMs by \citet{pokharel2014supervised,pokharel2016gaussian}. 
These data represent the results of an experiment designed to study the spread of the disease amongst 520 pepper plants raised in a greenhouse. Plants were evenly distributed across a 10$\times$26 meter area as shown in Figure~\ref{TSWV}. The experiment began on May 26, 1993 and finished on August 16, 1993. Plants were checked for the disease every 14 days, and ultimately 327 were infected. Following \citet{pokharel2014supervised,pokharel2016gaussian} these observation points are recorded to $t = 1, 2, \dots, 7$. We set the initial infection time to $t$ = 2 in line with the original data set. 

We here analyze the epidemic under two data availability scenarios. First, we assume that the event times of the TSWV disease are fully observed. Here, the infectious period was fixed at three time points (42 days) following \citet{pokharel2014supervised,pokharel2016gaussian}. Additionally, the last observed time point was at $t = 7$. Second, we assume the epidemic is partially observed. Specifically, we assume that the infection and removal times are unknown, and treat the reported infection times as the notified time points. This entails considerable uncertainty and makes the MCMC analysis much more time consuming (more than 13 times longer than the computation time of the first analysis), because it is necessary to estimate both incubation and delay periods along with the infection and removal times. 

The data is stored in the data file \code{tswv}, available in the \pkg{EpiILMCT} package. It contains a list of the TSWV epidemic data set for the two compartmental frameworks ($\mathcal{SIR}$ and $\mathcal{SINR}$) structured as a \code{`datagen'} class.

The following code shows how the TSWV data set can be extracted and the associated Euclidean distance matrix built. 
\begin{Sinput}
R> data("tswv", package = "EpiILMCT")
R> names(tswv)
\end{Sinput}
\begin{Soutput}
[1] "tswvsir"   "tswvsinr"
\end{Soutput}
\begin{Sinput}
R> plot(tswv$tswvsir$location, col = "gray", pch = 19)
R> k1 <- sum(tswv$tswvsir$epidat[,2] != Inf)
R> points(tswv$tswvsir$location[tswv$tswvsir$epidat[1:k1, 1], ], col = "red", 
+    pch = 19)
\end{Sinput}

Following \citet{pokharel2014supervised,pokharel2016gaussian}, we implement the distance-based continuous time ILM with power-law kernel and without susceptibility and transmissibility covariates. For the first analysis, an $\mathcal{SIR}$ distance-based continuous time ILM is used where the infectivity rate given in Equation~\ref{ratesir} becomes:
\[
\lambda_{j}(t) = \left(\alpha \sum_{i \in \mathcal{I}(t)}{d_{ij}^{-\beta}} \right) , \hspace{1cm} \alpha, \beta>0.
\]

To perform the MCMC, the \code{epictmcmc} function should be used with \code{datatype} set to \code{"known epidemic"}. Here, we assume exponential prior distributions with rate 0.01 for the model parameters $\alpha$ and $\beta$; and we request 150,000 MCMC samples. The code to achieve this is as follows: 

\begin{Sinput}
R> covsus <- list(NULL)
R> covsus[[1]] <- list(0.02, c("gamma", 1, 0.01, 0.01))
R> covsus[[2]] <- rep(1, length(tswv$tswvsir$epidat[,1]))
R> kernel1 <- list(2, c("gamma", 1, 0.01, 0.1))
\end{Sinput}
\begin{Sinput}
R> set.seed(524837)
R> tswv.full.observed <- epictmcmc(object = tswv$tswvsir, distancekernel = 
+    "powerlaw", datatype = "known epidemic", nsim = 150000, control.sus = 
+    covsus, kernel.par = kernel1)
R> plot(tswv.full.observed, plottype = "parameter", start = 10000, thin = 10, 
+    density = FALSE)
\end{Sinput}

\begin{figure}[!h]
\begin{center}
\includegraphics[width=\textwidth,height=6cm]{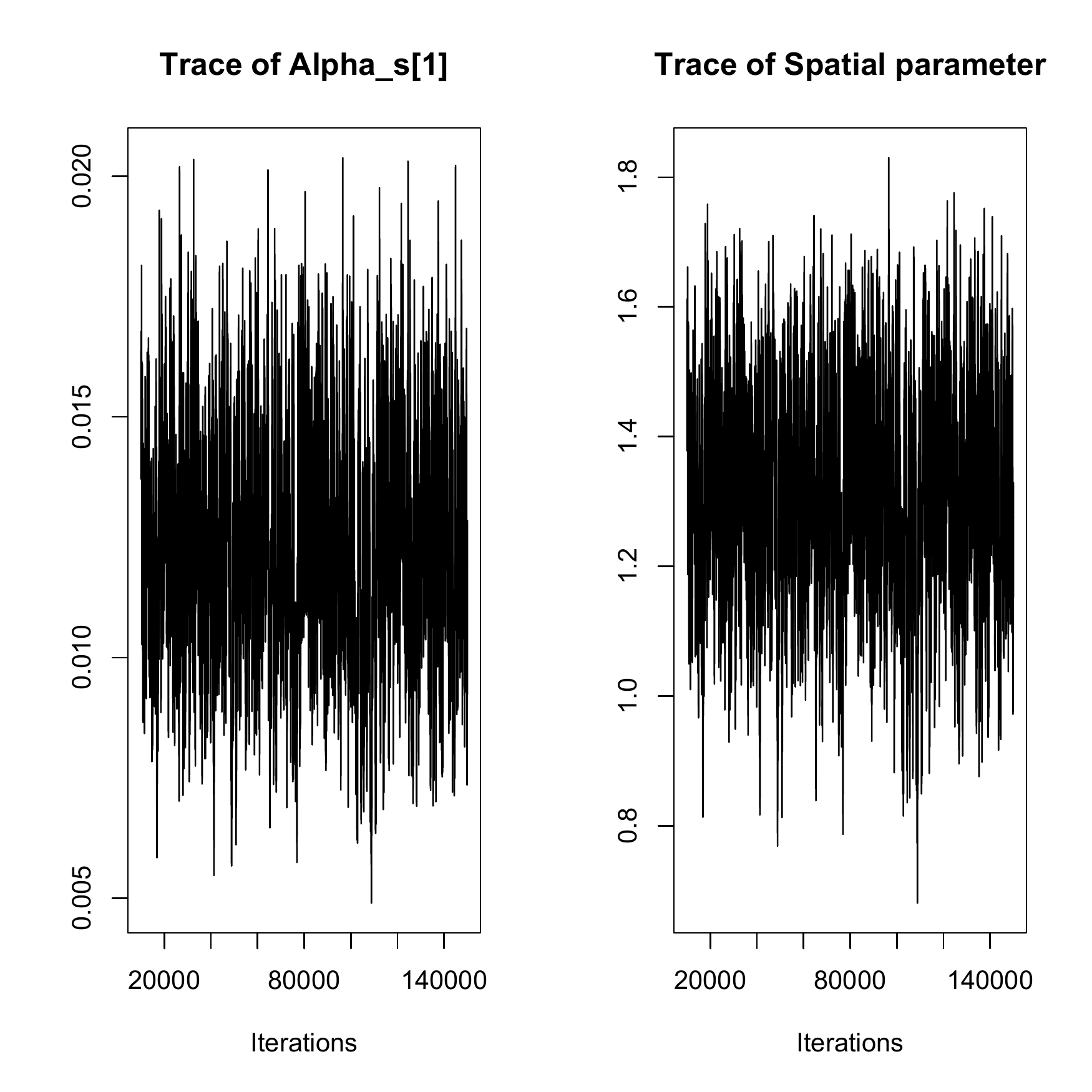}
\caption{The MCMC chains of the posterior distributions of the model parameter $\alpha$ and $\beta$ for fitting the fully observed TSWV data using the SIR distance-based continuous time ILM.}
\label{tswv-known-epidemic}
\end{center}
\end{figure}

Figure~\ref{tswv-known-epidemic} shows the resulting MCMC chains for the model parameters with a burn-in of 10,000 iterations and thinning interval of 10. The posterior mean of $\alpha$ and $\beta$ were $\hat{\alpha} = 0.012$ and $\hat{\beta} = 1.306$, with 95\% credible intervals of (0.007,0.017) and (0.973,1.592), respectively. The estimates of $\hat{\alpha}$ and $\hat{\beta}$ are consistent with those of \citet{pokharel2014supervised,pokharel2016gaussian}. The above \code{epictmcmc} function had a run time of one hour on an Apple MacBook Pro with i5-core Intel 2.4 GHz processors with 8 GB of RAM.

In the second analysis (i.e., where infection and removal times are treated as unknown), we assume notified times were observed for all infected individuals. Consequently, an $\mathcal{SINR}$ distance-based continuous time ILM is used where the infectivity rate given in Equation~\ref{ratesinr} becomes: 
\[
\lambda_{j}(t) = \left(\alpha \sum_{i \in \mathcal{N}^{-}(t)}{d_{ij}^{-\beta}} \right) +  \gamma \left(\alpha \sum_{i \in \mathcal{N}^{+}(t)}{d_{ij}^{-\beta}} \right).
\]

We here assume the risk of infection does not reduce after notification, and set the notification effect parameter to $\gamma$ = 1. The infectious period here is divided into the incubation and delay periods. We assume the total infectious period to be within three time points (42 days) following \citet{pokharel2014supervised,pokharel2016gaussian,brown2005tospoviruses}. Thus, we assumed the incubation periods to follow an exponential distribution such that $D_{i}^{(inc)} \sim$ Exp($\delta^{(inc)}$) with initial value of $\delta^{(inc)}$ = 1, whereas the delay periods is assumed to follow a gamma distribution such that $D_{i}^{(delay)} \sim \Gamma(10,\delta^{(delay)})$ with initial value of $\delta^{(delay)}$ = 5. We assign gamma prior distributions for both rates such that $\delta^{(inc)} \sim \Gamma(10, 10)$ and $\delta^{(delay)} \sim \Gamma(60, 12)$.  These choices are to cover the support of our assumptions about the infectious periods. For simplicity, we assume the infection time of the first infected plant is known. We set its incubation period to one time point.

We assign exponential prior distributions with rate 0.01 to the model parameters $\alpha$ and $\beta$. To perform the MCMC, we use the \code{epictmcmc} function with \code{type} and \code{datatype} set to \code{"SINR"} and \code{"unknown removal"}, respectively. At each iteration, the infection and removal times are updated in blocks of 10 randomly selected individuals (\code{blockupdate}). 
For faster implementation, we run the \code{epictmcmc} function in parallel to obtain 50,000 samples from four MCMC chains with four different sets of initial values of the model parameters and seed values. To do so, we set the argument \code{nchains = 4} and set \code{parallel = TRUE}. The number of cores to be used depends on the minimum number of the available cores and the number of chains (\code{nchains}). The following code was run using the four available cores of an Apple iMac with i5-core Intel 2.4 GHz processors and 8 GB of RAM.

\begin{Sinput}
R> covsus <- list(NULL)
R> covsus[[1]] <- list(NULL)
R> covsus[[1]][[1]] <- c(0.02, 0.1, 1.5, 3)
R> covsus[[1]][[2]] <- c("gamma", 1, 0.01, 0.01)
R> covsus[[2]] <- rep(1, length(tswv$tswvsir$epidat[,1]))
R> kernel1 <- list(c(0.1, 5, 1, 10), c("gamma", 1, 0.01, 0.1))
R> delta1 <- list(NULL)
R> delta1[[1]] <- c(1,10)
R> delta1[[2]] <- matrix(c(10, 5, 1, 0.5, 15, 2, 1, 15), ncol = 4, nrow = 2)
R> delta1[[3]] <- matrix(c(10, 60, 10, 12), ncol = 2, nrow = 2)
\end{Sinput}
\begin{Sinput}
R> set.seed(524837)
R> tswv.unknown.remov.infect1 <- epictmcmc(object = tswv$tswvsinr, 
+    distancekernel = "powerlaw", datatype = "unknown removal", blockupdate = 
+    c(1, 10), nsim = 50000, nchains = 4, parallel = TRUE, control.sus = 
+    covsus, kernel.par = kernel1, delta = delta1)
\end{Sinput}

\begin{figure}[!h]
\begin{center}
\includegraphics[width=\textwidth,height=12cm]{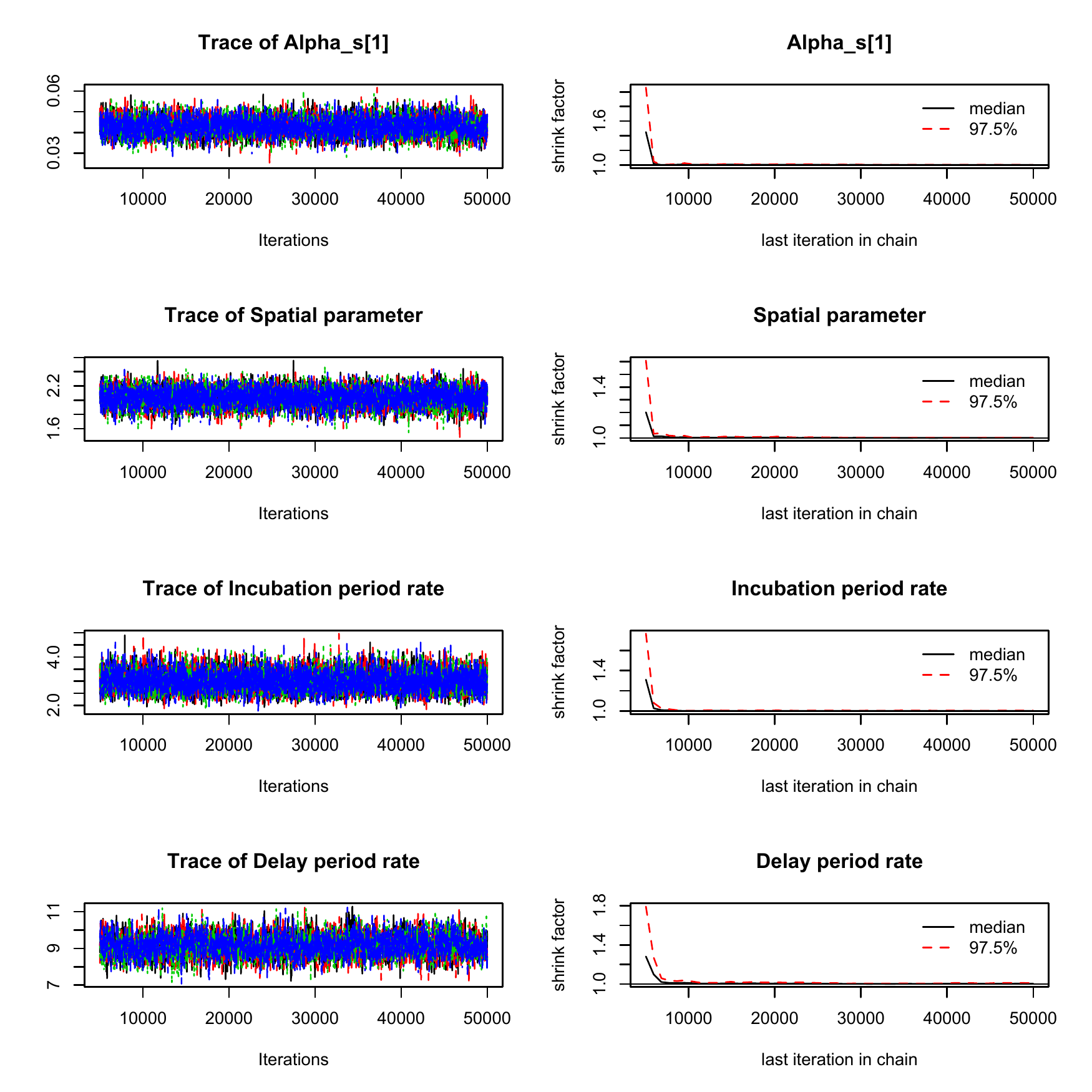}
\caption{The four MCMC chains (left) and Gelman-Rubin convergence diagnostic (right) plots of the posterior distributions of the model parameters $\alpha$, $\beta$, $\delta^{(inc)}$ and $\delta^{(delay)}$ for fitting the partially observed TSWV data (unknown infection and removal times) using the SINR distance-based continuous time ILM.}
\label{tswv-unknown-epidemic}
\end{center}
\end{figure}

Figure~\ref{tswv-unknown-epidemic} shows the MCMC trace plots and Gelman-Rubin convergence diagnostic plots for the model parameters $\alpha$, $\beta$, $\delta^{(inc)}$ and $\delta^{(delay)}$ with a burn-in of 5,000 iterations removed and a thinning interval of 10 for the four MCMC chains. Figure~\ref{tswv-unknown-epidemic} can be produced using the following code:
\begin{Sinput}
R> layout(matrix(c(5, 1, 6, 2, 7, 3, 8, 4), ncol = 2, byrow = T))
R> m1 <- window(tswv.unknown.remov.infect1$parameter.samples[[1]], start = 
+    5000)
R> m2 <- window(tswv.unknown.remov.infect1$parameter.samples[[2]], start = 
+    5000)
R> m3 <- window(tswv.unknown.remov.infect1$parameter.samples[[3]], start = 
+    5000)
R> m4 <- window(tswv.unknown.remov.infect1$parameter.samples[[4]], start = 
+    5000)
R> gelman.plot(mcmc.list(m1, m2, m3, m4), auto.layout = FALSE)
R> plot(tswv.unknown.remov.infect1, plottype = "parameter", start = 5000, 
+    thin = 10, density = FALSE, smooth = FALSE, auto.layout = FALSE)
\end{Sinput}

The posterior means and 95\% credible intervals of these parameters are given in Table~\ref{unknown-tswv-table}. The MCMC chains show good mixing with both trace and Gelman-Rubin plots suggesting convergence. 

\begin{figure}[!h]
\begin{center}
\includegraphics[width=0.9\textwidth,height=9cm]{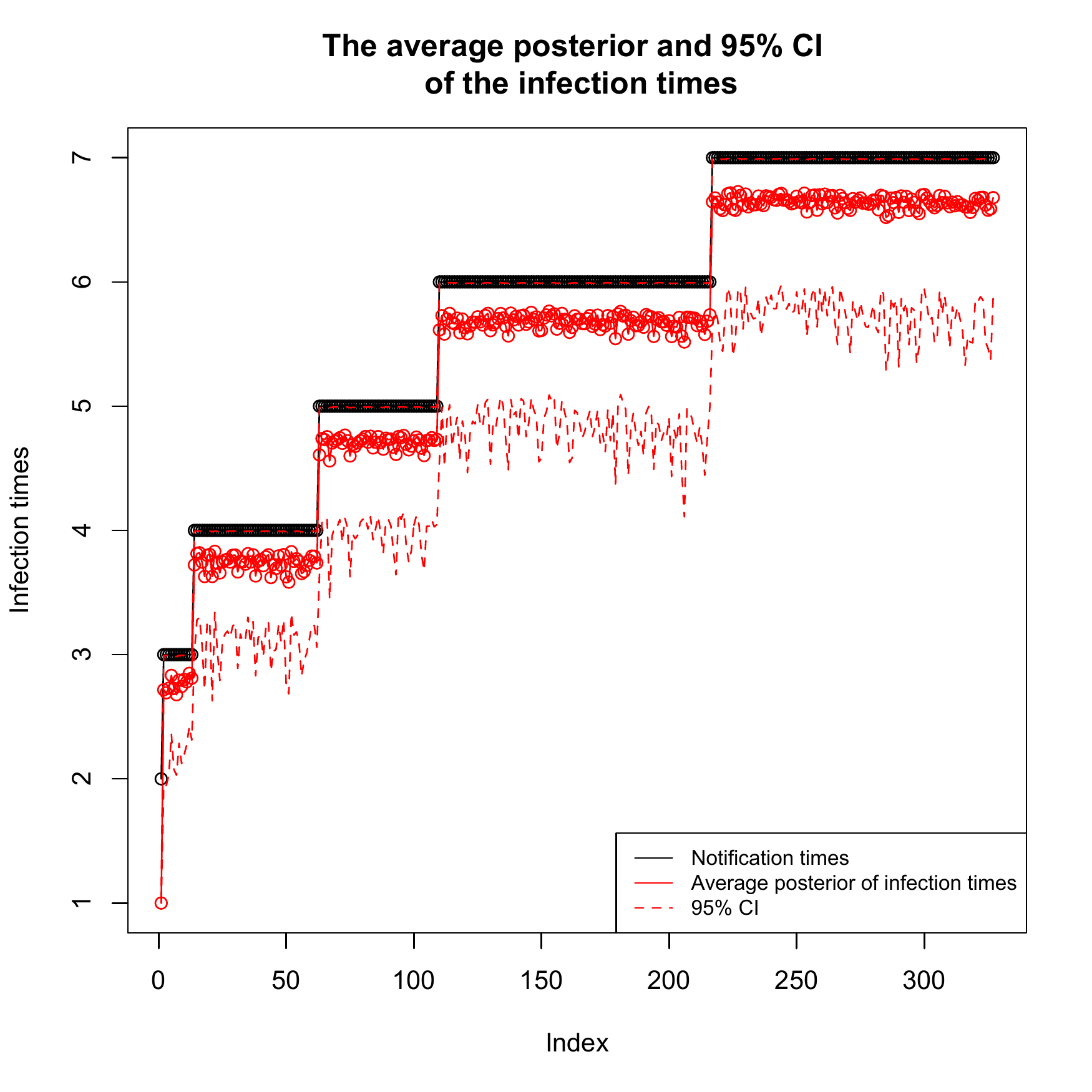}
\caption{The posterior means (solid red line) and 95$\%$ credible intervals (dotted red lines) of the infection times for fitting the partially observed TSWV data (unknown infection and removal times) using the SINR distance-based continuous time ILM. The black line represents the observed notification times.}
\label{infection-tswv}
\end{center}
\end{figure}

\begin{figure}[!h]
\begin{center}
\includegraphics[width=0.9\textwidth,height=9cm]{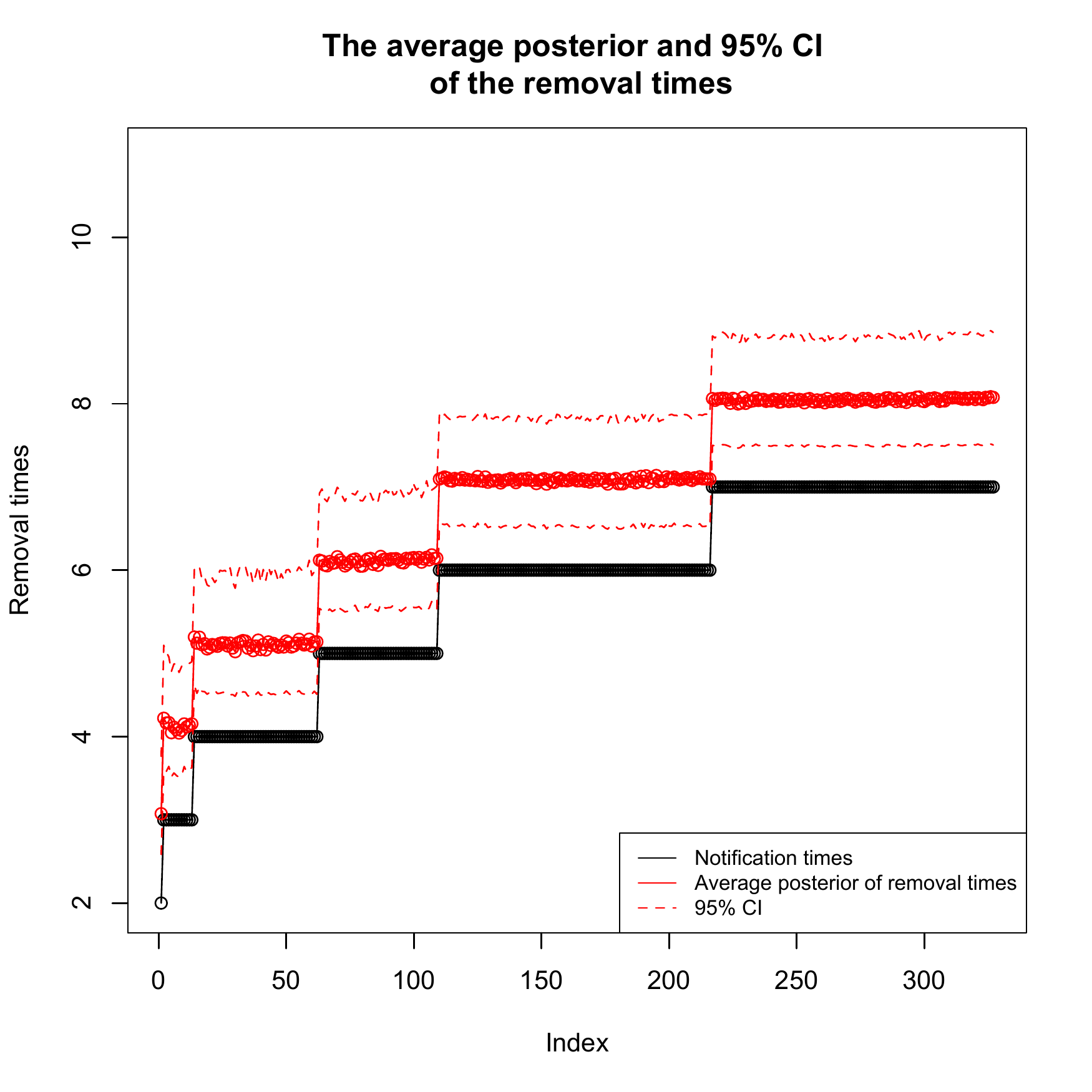}
\caption{The posterior means (solid red line) and 95$\%$ credible intervals (dotted red lines) of the removal times for fitting the partially observed TSWV data (unknown infection and removal times) using the SINR distance-based continuous time ILM. The black line represents the observed notification times.}
\label{removal-tswv}
\end{center}
\end{figure}

Figures~\ref{infection-tswv} and \ref{removal-tswv} show the posterior means and 95\% credible intervals of the infection and removal times. Theses figures can be produced using the following commands:
\begin{Sinput}
R> plot(tswv.unknown.remov.infect1, epi = tswv$tswvsinr, plottype = 
+    "inf.times", start = 5000, thin = 10)
R> plot(tswv.unknown.remov.infect1, epi = tswv$tswvsinr, plottype = 
+    "rem.times", start = 5000, thin = 10)
\end{Sinput}

Using the \code{summary} function of the object \code{tswv.unknown.remov.infect1}, the posterior means (95\% CIs) of the incubation and delay periods were found to be 0.320 (0.242,0.414) and 1.082 (0.957,1.224) observation time points, respectively, indicating an average infectious period of 19.628 days (1.402 time points).

Note that, infection and removal times are updated here in blocks of 10 (via the \code{blockupdate} argument) for reasons of computational efficiency.
The \code{epictmcmc} function had a run time of 9.51 hours using the parallel method with 4 cores, but this was computationally much more efficient than if single updates were used ($\approx$ 124 hours, calculated based on ten MCMC iterations).

\begin{table}[!h]
\begin{center} 
\begin{tabular}{c|cccc}
\hline
		&$\alpha$		&$\beta$&	$\delta^{(inc)}$ & $\delta^{(delay)}$	\\	
\hline		
Mean	&	0.043	&	2.037	&	2.992 	&	9.139 	\\
95\% CI 	&	(0.034, 0.051)	&	(1.780, 2.275)	&	(2.264, 3.874)	&	(8.046, 10.292)	\\
\hline		

\end{tabular}
\end{center}
\caption{The posterior means and 95\% credible intervals (CIs) of the model parameters, with a burn-in of 5,000 iterations and thinning interval of 10 for each of the four MCMC chains, for fitting the TSWV using the SINR distance-based continuous time ILM under the assumption of unknown removal and infection times.}
\label{unknown-tswv-table}
\end{table}%

\section{Conclusion}

This paper introduces the \proglang{R} software package \pkg{EpiILMCT}, which facilitates the use of a broad range of continuous time ILMs under two compartmental frameworks ($\mathcal{SIR}$ and $\mathcal{SINR}$). It also allows for the analysis of partially observed infectious diseases data, achieved using data augmented MCMC within a Bayesian framework. We illustrated the package by fitting continuous time ILMs on simulated and real epidemic data. The paper did not cover all functionality of the package. For instance, we did not illustrate incorporating both distance and network in the kernel function, or allowing for nonlinearity between the susceptibility and transmissibility risk factors and the infection rate. However, implementation of such facets is simple. Additional functionality that was not covered in Sections~5 can be found via \code{help(package = "EpiILMCT")}.

Also, it is possible to use \pkg{EpiILMCT} to test the efficacy of disease control strategies (eg. vaccination or culling) via simulation study. This can be done by simulating epidemics in small time steps and then manipulating infection and/or removal times according to a given control policy, before simulating the next step of the epidemic simulation conditional upon the manipulated epidemic history just determined. We illustrate this via a simple ring-culling strategy in Appendix~\ref{appendix.3}.

In terms of future developments, the authors intend to expand the modelling framework to allow for latent periods (i.e., susceptible-exposed-infectious-removed ($\mathcal{SEIR}$) and susceptible-exposed-infectious-notified-removed ($\mathcal{SEINR}$)). This would be useful for many disease systems in which the time between infection (exposure) and infectiousness cannot be reasonably ignored. Additionally, expanding the compartmental frameworks to allow for reinfection would also be useful for diseases such as influenza. That is, we could allow for frameworks: susceptible-infectious-susceptible ($\mathcal{SIS}$), susceptible-exposed-infectious-susceptible ($\mathcal{SEIS}$), etc.    

Another development could involve incorporating more data uncertainty into the analyses, especially under the network-based model, is an option for future development of this package \pkg{EpiILMCT}. For example, networks are often only partially observed. However, the data augmentation could easily make the computation time for data analyses prohibitive. Various strategies for mitigating this might be available. 
For example, approximate forms of inference such as Gaussian process emulation \citep{pokharel2016gaussian}, approximate Bayesian computation \citep{beaumont2009adaptive}, machine learning based model classification \citep{pokharel2014supervised}, data-sampled likelihood approximation \citep{malik2016parameterizing}, or data-aggregation \citep{deeth2016spatial} could all prove useful for overcoming these computational issues. Finally, it would be possible to extend our modelling framework to allow for multiple, interacting disease strains or pathogens \citep{romanescu2016modeling}.  

\section*{Acknowledgments}

This work was funded by the Ontario Ministry of Agriculture, Food and Rural Affairs (OMAFRA), and the Natural Sciences and Engineering Research Council of Canada (NSERC). Almutiry was also funded by Qassim University through the Saudi Arabian Cultural Bureau in Canada. Warriyar was funded by the University of Calgary Eyes High Post Doctoral Scholarship scheme. We thank the editor and referees for their valuable suggestions and comments, which greatly improved both the software package and this manuscript.

\bibliography{EpiILMCT}

\newpage
\appendix

\section[A.]{The likelihood function of the general $\mathcal{SINR}$ continuous time ILMs:}\label{appendix.2}

{\small
\begin{eqnarray}
L(\boldsymbol{I},\boldsymbol{N},\boldsymbol{R}|\boldsymbol{\theta}) &=& 
\prod_{j=2}^{m}{\left(\epsilon +\sum_{i:I_{i} < I_{j} \leq N_{i}}{\lambda_{ij}^{-}(I_{j})} + \sum_{i:N_{i} < I_{j} \leq R_{i}}{\lambda_{ij}^{+}(I_{j})} \right)} \nonumber \\
&\times& \exp \left\{ -\int_{I_{1}}^{t_{obs}}{\left(\sum_{i \in \mathcal{S}(u)}{\epsilon}  +  \sum_{i \in \mathcal{I}(u)}\sum_{j \in \mathcal{S}(u)}{\lambda_{ij}^{-}(u-I_{i})}  +  \sum_{i \in \mathcal{N}(u)}\sum_{j \in \mathcal{S}(u)}{ \lambda_{ij}^{+}(u-I_{i})}  \right) du} \right\} \nonumber \\
&\times& \prod_{i=1}^{m}{f(\mathcal{D}^{(inc)}_{i};\delta^{(inc)})} \prod_{i=1}^{m}{f(\mathcal{D}^{(delay)}_{i};\delta^{(delay)})} \nonumber \\ 
&=& 
\prod_{j=2}^{m}{\left(\epsilon +\sum_{i:I_{i} < I_{j} \leq N_{i}}{\lambda_{ij}^{-}(I_{j})}+\sum_{i:N_{i} < I_{j} \leq R_{i}}{\lambda_{ij}^{+}(I_{j})}\right)} \nonumber \\
&\times& \exp \left\{ -\sum_{i=1}^{m}{      \left(\sum_{j=1}^{N}{( (t_{obs} \wedge N_{i} \wedge I_{j}) - (I_{i} \wedge I_{j})) \lambda_{ij}^{-}(I_{j}) } \right)} \right\} \nonumber \\
&\times& \exp \left\{- \sum_{i=1}^{m}{\left(\sum_{j=1}^{N}{( (t_{obs} \wedge R_{i} \wedge I_{j}) - (I_{i} \wedge I_{j})) - ( (t_{obs} \wedge N_{i} \wedge I_{j}) - (I_{i} \wedge I_{j})) \lambda_{ij}^{+}(I_{j})}\right)} \right\} \nonumber \\
&\times& \exp \left(- \epsilon \sum_{i=1}^{N}{\left[(t_{obs} \wedge I_{i}) - I_{1}\right]} \right) \nonumber \\
&\times& \prod_{i=1}^{m}{f(\mathcal{D}^{(inc)}_{i};\delta^{(inc)})}~\prod_{i=1}^{m}{f(\mathcal{D}^{(delay)}_{i};\delta^{(delay)})} \hspace{2cm} \delta^{(inc)}, \delta^{(delay)} > 0,
	 \label{eq:eqsinr}
\end{eqnarray}
}

\noindent where the wedge symbol $\wedge$ denotes the minimum operator; and $\mathcal{D}^{inc}_{i}$ and $\mathcal{D}^{delay}_{i}$ are the incubation and delay periods such that $\mathcal{D}^{inc}_{i} = N_{i} - I_{i}$ and  $\mathcal{D}^{delay}_{i} = R_{i} - N_{i}$, respectively. 

\newpage
\section[B.]{\proglang{R} code for extracting individual level data from \pkg{surveillance}}\label{appendix.1}

Here, we illustrate the extraction of individual level data from the \pkg{surveillance} package for use in the \pkg{EpiILMCT} package. We consider the toy data set (\code{fooepidata}) representing a population of 100 individuals that is used in the \code{twinSIR} examples of the \pkg{surveillance} package \citep{surveillance}.
 
The data can be found in \pkg{surveillance} via \code{data("fooepidata", package = "surveillance")}.  
\begin{Sinput}
R> library("surveillance")
R> data("fooepidata")
R> names(fooepidata)
\end{Sinput}
\begin{Soutput}
 [1] "BLOCK"   "id"      "start"   "stop"    "atRiskY" "event"   "Revent" 
 [8] "x"       "y"       "z1"      "z2"      "B1"      "B2"     
\end{Soutput}
The \code{fooepidata} event history consists of 178 time \code{BLOCK}s of 100 rows, where each row describes the state of individual id during the corresponding time interval (\code{start}, \code{stop}).
\begin{Sinput}
R> head(fooepidata, n = 5)
\end{Sinput}
\begin{Soutput}
     BLOCK id start      stop atRiskY event Revent            x          y
1        1  1     0 0.6970682       1     0      0  1.262954285  0.7818592
246      1  2     0 0.6970682       1     0      0 -0.326233361 -0.7767766
369      1  3     0 0.6970682       1     0      0  1.329799263 -0.6159899
612      1  4     0 0.6970682       1     0      0  1.272429321  0.0465803
760      1  5     0 0.6970682       1     0      0  0.414641434 -1.1303858
     z1        z2 B1 B2
1     0 0.0000000  0  0
246   1 0.6931472  0  0
369   0 1.0986123  0  0
612   1 1.3862944  0  0
760   1 1.6094379  0  0
[....]
\end{Soutput}

The \code{start} and \code{stop} variables represent the start and end of interval time points (in continuous time) that indicate the waiting time between consequence event times (infection and removal times). The binary variables \code{event} and \code{Revent} are used to indicate the occurrence of newly infected or removed individuals at the stop time of each time interval (\code{BLOCK}), respectively. 
Thus, the \code{stop} time is taken to be the infection or removal times of the infected or removed individuals in each time interval. 
The coordinates of individuals is represented in columns \code{x} and \code{y}. The 
\code{fooepidata} data set contains also endemic and epidemic covariates. Endemic covariates are represented by the columns named \code{z1} and \code{z2} (the exact interpretation of these covariates is not given). Epidemic covariates are represented by the columns named \code{B1} and \code{B2}, and they indicate the count of currently infective individuals for each individual within, and greater than one unit distance, respectively. See (\code{help(epidata, package= "surveillance")}) for more details about the data structure. From this data set,  we extract only the event times and XY coordinates of each individual, ignoring the purely spatial epidemic covariates which are directly modelled by the distance kernel in \pkg{EpiILMCT}.
\begin{Sinput}
R> epi <- summary(fooepidata)$byID
R> loc <- summary(fooepidata)$coordinates
R> epi[is.na(epi)] <- Inf
R> epi <- transform(epi, period = ifelse(is.infinite(time.I), 0, time.R - 
+    time.I))
R> epi$id <- as.integer(as.character(epi$id))
R> epidat <- as.matrix(epi[c("id", "time.R", "period", "time.I")])
R> library("EpiILMCT")
R> epi <- as.epidat(type = "SIR", kerneltype = "distance", inf.time = 
+    epidat[, 4], rem.time = epidat[, 2], id.individual = epidat[, 1], 
+    location = loc)
\end{Sinput}

The object $epi$ of class \code{`datagen'} can be now used in the \pkg{EpiILMCT} package using the model given in Equation~\ref{ratesir} without covariates through the following code: 
\begin{Sinput}
R> set.seed(101)
R> sus.par <- list(NULL)
R> sus.par[[1]] <- list(0.1, c("gamma", 1, 0.001, 0.005))
R> sus.par[[2]] <- matrix(rep(1, length(epi$epidat[, 1])), ncol = 1)
R> kernel <- list(0.1, c("gamma", 1, 0.001, 0.1))
R> spark <- list(0.1, c("gamma", 1, 0.001, 0.05))
R> mcmc1 <- epictmcmc(object = epi, distancekernel = "powerlaw", datatype = 
+    "known epidemic", nsim = 50000, control.sus = sus.par, kernel.par = 
+    kernel, spark.par = spark)
\end{Sinput}
We include the spark term here to best model the endemic component used in the \code{twinSIR} model. 
The inclusion of the spark term also allows for the fact that there are no infectious individuals during times intervals (\code{BLOCK}) of the epidemic.
The infection of individuals in these periods is captured by the endemic part in \code{twinSIR} function. 

Without incorporating the spark term in the \code{epictmcmc} function, a zero likelihood will result, preventing the successful fitting of the model to the data. To get the output estimates of the model parameters, we used the \proglang{S3} method \code{summary.epictmcmc} as follows:
\begin{Sinput}
R> summary(mcmc1, start = 1000)
\end{Sinput}
\begin{Soutput}
********************************************************* 
Model: SIR distance-based continuous-time ILM 
Method: Markov chain Monte Carlo (MCMC) 
Data assumption: fully observed epidemic 
number.chains : 1 chains 
number.iteration : 49000 iterations 
number.parameter : 3 parameters 
********************************************************* 
 1. Empirical mean and standard deviation for each variable,
plus standard error of the mean:
                        Mean         SD    Naive SE Time-series SE
Alpha_s[1]        0.00889042 0.00110553 4.99425e-06    1.54972e-05
Spark             0.00778819 0.00436839 1.97342e-05    6.85098e-05
Spatial parameter 0.94175614 0.18258926 8.24846e-04    3.82173e-03
 2. Quantiles for each variable:
                        2.5%        25%        50%        75%     97.5%
Alpha_s[1]        0.00686125 0.00811098 0.00884201 0.00962789 0.0111386
Spark             0.00131269 0.00452088 0.00718377 0.01034424 0.0180864
Spatial parameter 0.54032375 0.82833931 0.95593444 1.07000331 1.2615483
 3. Empirical mean, standard deviation, and quantiles for the log likelihood,
          Mean             SD       Naive SE Time-series SE 
  -2.30176e+02    1.23456e+00    5.57714e-03    2.00104e-02 
    2.5%      25%      50%      75%    97.5% 
-233.367 -230.757 -229.854 -229.263 -228.752 
 4. acceptance.rate : 
       Alpha_s[1]             Spark Spatial parameter 
         0.253945          0.169543          0.810156 
\end{Soutput}

We also demonstrate the modelling of these data using \code{twinSIR} function with no endemic covariates. However, a baseline term (baseline hazard rate) will be included in this case in the endemic component to represent the background rate of infection in the population, as explained in the note Section in \code{help( twinSIR, package = "surveillance" )}. The following code illustrates the use of \code{twinSIR} in analyzing this data set.
\begin{Sinput}
R> fit1 <- twinSIR( ~ B1 + B2, data = fooepidata)
R> summary(fit1)
\end{Sinput}
\begin{Soutput}
Call:
twinSIR(formula = ~B1 + B2, data = fooepidata)
Coefficients:
                  Estimate Std. Error z value Pr(>|z|)    
B1                0.023960   0.004208   5.693 1.25e-08 ***
B2                0.003395   0.001119   3.034  0.00241 ** 
cox(logbaseline) -6.010580   0.659257  -9.117  < 2e-16 ***
---
Signif. codes:  0 '***' 0.001 '**' 0.01 '*' 0.05 '.' 0.1 ' ' 1
Total number of infections:  88 
One-sided AIC: 474.05
Log-likelihood: -235.2
Number of log-likelihood evaluations: 26 
\end{Soutput}

\begin{figure}[t]
\begin{center}
\includegraphics[width=0.8\textwidth,height=9cm]{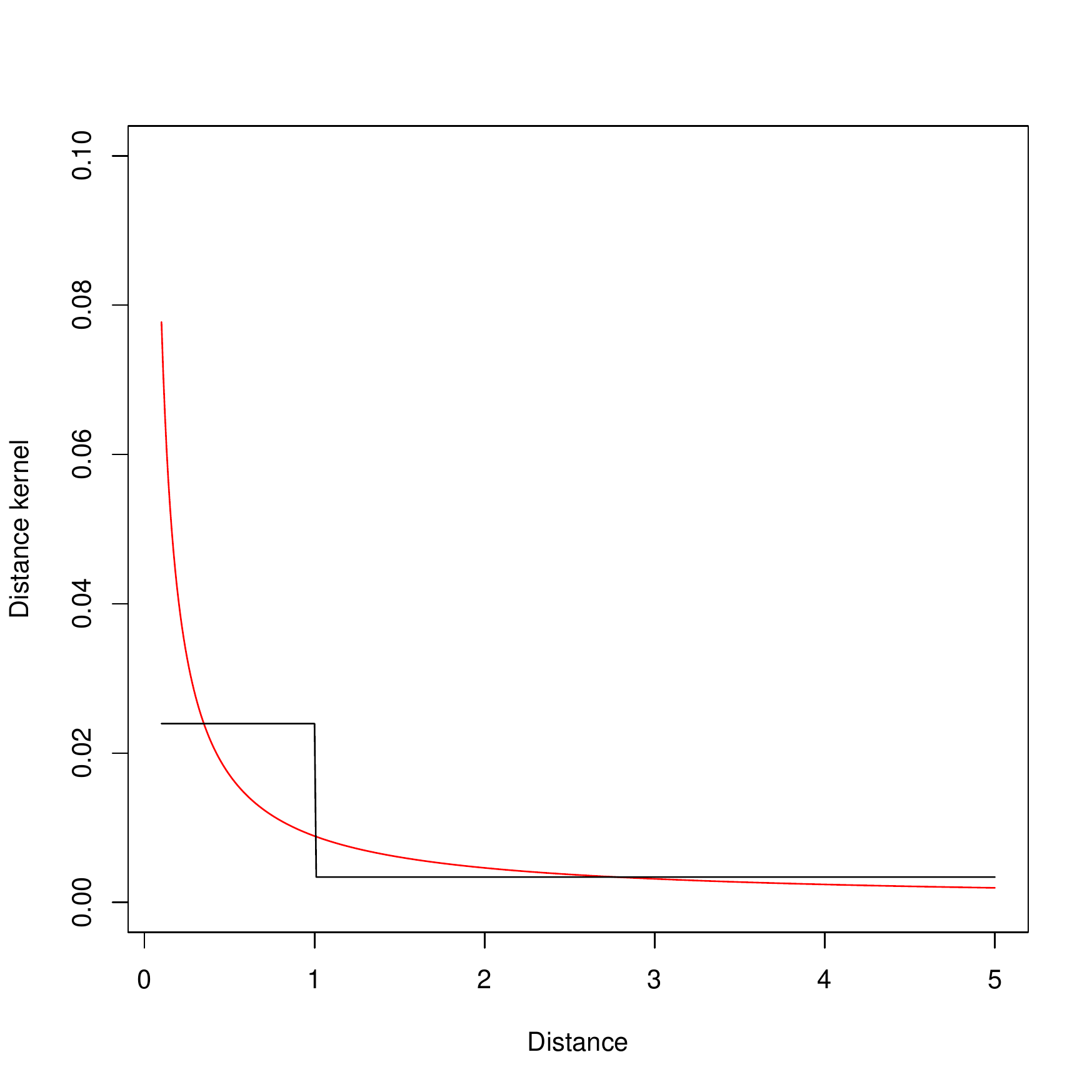}
\caption{The marginal posterior distribution of the distance kernels. Black line represent the spatial terms of the \pkg{surveillance} package, and red line represents the distance kernel function of the \pkg{EpiILMCT} package.}
\label{surveillance_EpILMCT}
\end{center}
\end{figure}

The posterior means of the ILM parameters ($\alpha$, $\beta$) are 0.009 and 0.945, respectively. Figure~\ref{surveillance_EpILMCT} shows the ILM power-law distance kernel function under the posterior mean, along with the distance function suggested by the MLEs of the model parameters from the \code{twinSIR} analysis. We can see broad agreement, although the step function assumption of the \code{twinSIR} seems less reasonable than the continuous decay of the ILM kernel for short distances (< 1 distance unit).

\newpage
\section[C.]{\proglang{R} code for implementing ring-based control strategy}\label{appendix.3}

Here, we illustrate the use of the \pkg{EpiILMCT} package in testing the efficacy of a ring-based control strategy for mitigating the spread of disease. We consider an example in which an infectious disease is transmitted between 625 individuals located in a square area of 50$\times$50 units. These individuals could be thought to represent farms or trees, say. We implement control measures upon all individuals within a circle of $r$ radius of newly infected individuals. This control strategy essentially places these individuals in the removed set. These measures could be thought to represent vaccination or quarantine, but here we assume it is a culling strategy.

To illustrate we first simulate the XY coordinates of individuals from a uniform distribution. This is done as follows:
\begin{Sinput}
R> library("EpiILMCT")
R> set.seed(101)
R> n <- 625
R> loc <- matrix(cbind(runif(n, 0, 50), runif(n, 0, 50)), ncol = 2, nrow = n)
\end{Sinput}

We assume that the epidemic starts with an initial infected individual $k = 386$, who has an infection time $I_{1} = 0$ and an infectious period of 3 days. We then implement the culling policy within an epidemic simulation study using the \code{datagen} command to simulate epidemics in a specified small time steps (e.g., a day at a time). This is done by setting the option \code{tmax}, and starting each new simulation step with initially infected and removed individuals set according to the epidemic history, and the culling policy implemented at the current time step. This is done using the \code{initialepi} option. We build a \code{control.strategy} function to implement the above culling policy using an $\mathcal{SIR}$ distance-based continuous time ILM with power-law kernel and no covariates, in which the infectivity rate given in Equation~\ref{ratesir} becomes:
\[
\lambda_{j}(t) = \left(\alpha \sum_{i \in \mathcal{I}(t)}{d_{ij}^{-\beta}} \right) , \hspace{1cm} \alpha, \beta>0,
\]
with infectious periods assumed to follow a gamma distribution such that $\gamma_{i} \sim \Gamma(6,\delta)$. 

\begin{Sinput}
   R> control.strategy <- function(init.epi, location, inf.time, par.sus,
   +    par.ker, delt, cov.sus = NULL, radius) {
   +    n <- length(location[, 1])
   +    tss <- init.epi
   +    cov1 <- cov.sus
   +    dis <- as.matrix(dist(location))
   +    for (i in 2:length(inf.time)) {
   +      mn <- sum(tss[, 4] <= inf.time[i-1])
   +      initial1 <- matrix(tss[1:mn,], ncol = 4, nrow = mn)
   +      tss1 <- datagen(type = "SIR", kerneltype = "distance",
   +        kernelmatrix = location, distancekernel = "powerlaw",
   +        initialepi = initial1, tmax = inf.time[i], suspar = par.sus,
   +        transpar = NULL, kernel.par = par.ker, delta = delt,
   +        transcov = NULL, suscov = cov1)
   +      tss <- tss1$epidat
   +      newlyinfected <- tss[which(tss[, 4] > inf.time[i-1] &
   +      tss[, 4] <= inf.time[i]), 1]
   +      num.infected  <- sum(tss[, 2] != Inf)
   +      uninfected <- tss[(num.infected+1):n, 1]
   + # All individuals within the ring of a radius (radius) of
   + # the newly infected are removed:
   +      for (j in 1:length(newlyinfected)) {
   +        mk <- as.integer(which(dis[newlyinfected[j], uninfected] <
   +          radius))
   +        if (length(mk) > 0) {
   +          cov1[uninfected[mk], ] = 0
   +        }
   +      }
   +    }
   +    list(tss1, cov1)
   +    }
\end{Sinput}

Let us assume we have estimates of the model parameters as $\hat{\alpha}$ = 1.5, $\hat{\beta}$ = 4, and $\hat{\delta}$ = 2. Using these estimates, we test the above function for eight values of the radius of the culling policy, and obtain 32 replicated epidemics for each radius setting. 
The code to achieve this is as follows:

\begin{Sinput}
   R> id.init <- 386
   R> inf.period.init <- 3
   R> kl <- which(seq_len(625) != id.init)
   R> init.epi <- rbind(c(386, inf.period.init, inf.period.init, 0),
   +    cbind(kl, rep(Inf, 624), rep(0, 624), rep(Inf, 624)))
   R> rr <- seq_len(8)
   R> inf.time <- seq(0, 30, by = 1)
   R> par.sus <- 1.5
   R> par.ker <- 4.0
   R> delt <- c(6, 2)
   R> sus.cov <- matrix(rep(1, 625), ncol = 1)
   R> ninfected <- matrix(0, ncol = 32, nrow = length(rr))
   R> numb.culled <- matrix(0, ncol = 32, nrow = length(rr))
   R> len.infection <- matrix(0, ncol = 32, nrow = length(rr))
   R> for (i in 1:length(rr)) {
   +    for (j in 1:32) {
   +      epi.cont <- control.strategy(init.epi, location = loc, inf.time,
   +        par.sus, par.ker, delt, cov.sus = sus.cov, radius = rr[i])
   +      ninfected[i, j] <- sum(epi.cont[[1]]$epidat[, 2] != Inf)
   +      numb.culled[i, j] <- n - apply(epi.cont[[2]], 2, sum)
   +      len.infection[i, j] <- max(epi.cont[[1]]$epidat[1:ninfected[i, j],
   +        2]) - min(epi.cont[[1]]$epidat[1:ninfected[i, j], 4])
   +    }
   +  }
\end{Sinput}

The output of the above loops is an $8 \times 32$ matrices of the number of infected and culled individuals and the length of epidemics for the radius set. We then use the function \code{apply} from the \pkg{base} package \citep{CRAN} to get the average of each summary at each radius, and plot them versus radius using the following code: 
\begin{Sinput}
R> plot(rr, apply(ninfected, 1, mean), type = "o", ylab = "Number of 
+    individuals", xlab = "radius", ylim = c(0, n), pch = 19)
R> lines(rr, apply(numb.culled, 1, mean), type = "o", pch = 19, col = "red")
R> legend("topright", c("Average number of infected individuals", "Average 
+    number of culled individuals"), col = c("black", "red"), lty = c(1, 1), 
+    pch = c(19, 19))
R> plot(rr, apply(len.infection, 1, mean), type = "o", ylab = "Length of 
+    epidemic", xlab = "radius", pch = 19)
\end{Sinput}

\begin{figure}[!ht]
\begin{center}
\includegraphics[width=\textwidth,height=8cm]{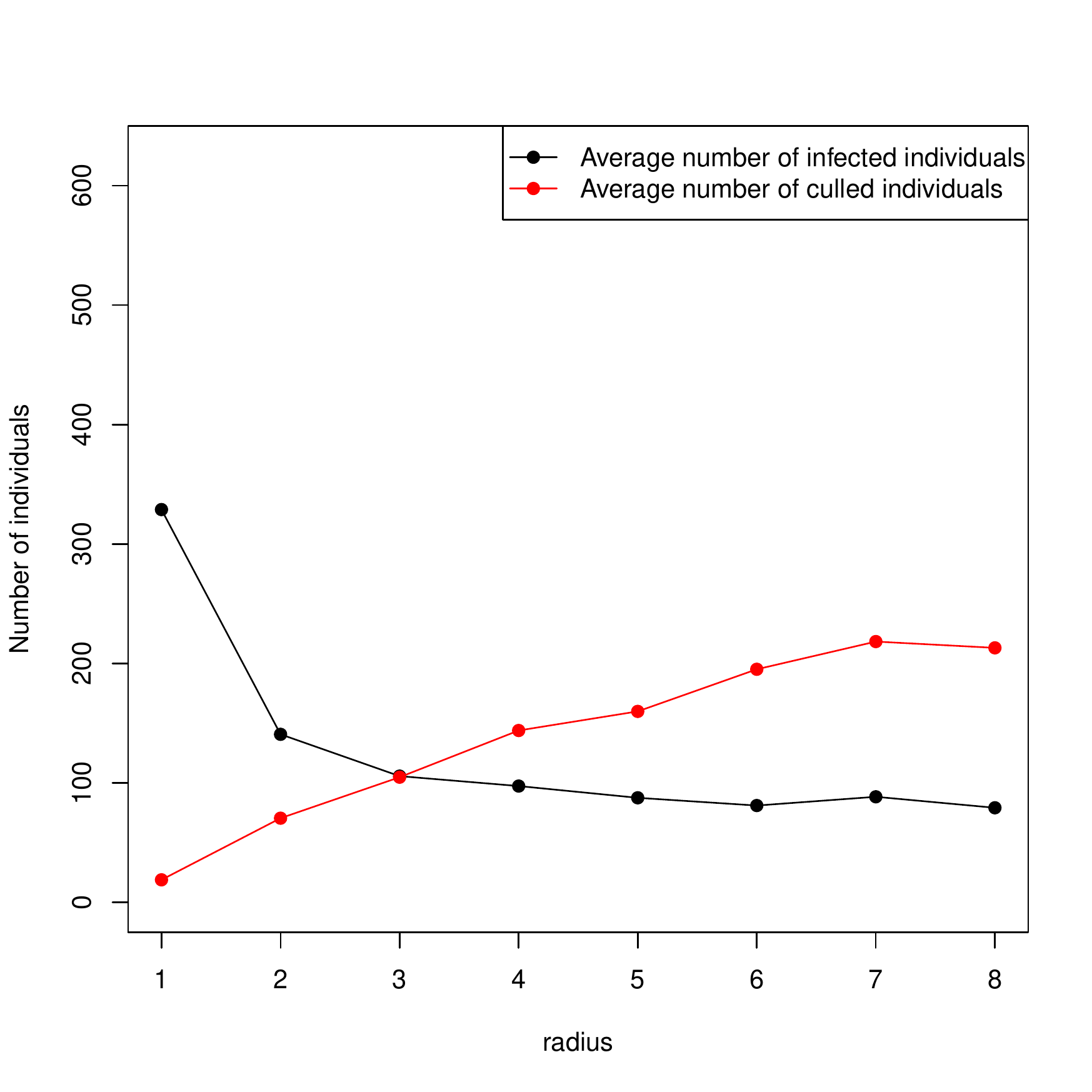}
\caption{The average number of infected (black) and culled (red) individuals at each radius.}
\label{infec-culled}
\end{center}
\end{figure}

\begin{figure}[!h]
\begin{center}
\includegraphics[width=\textwidth,height=8cm]{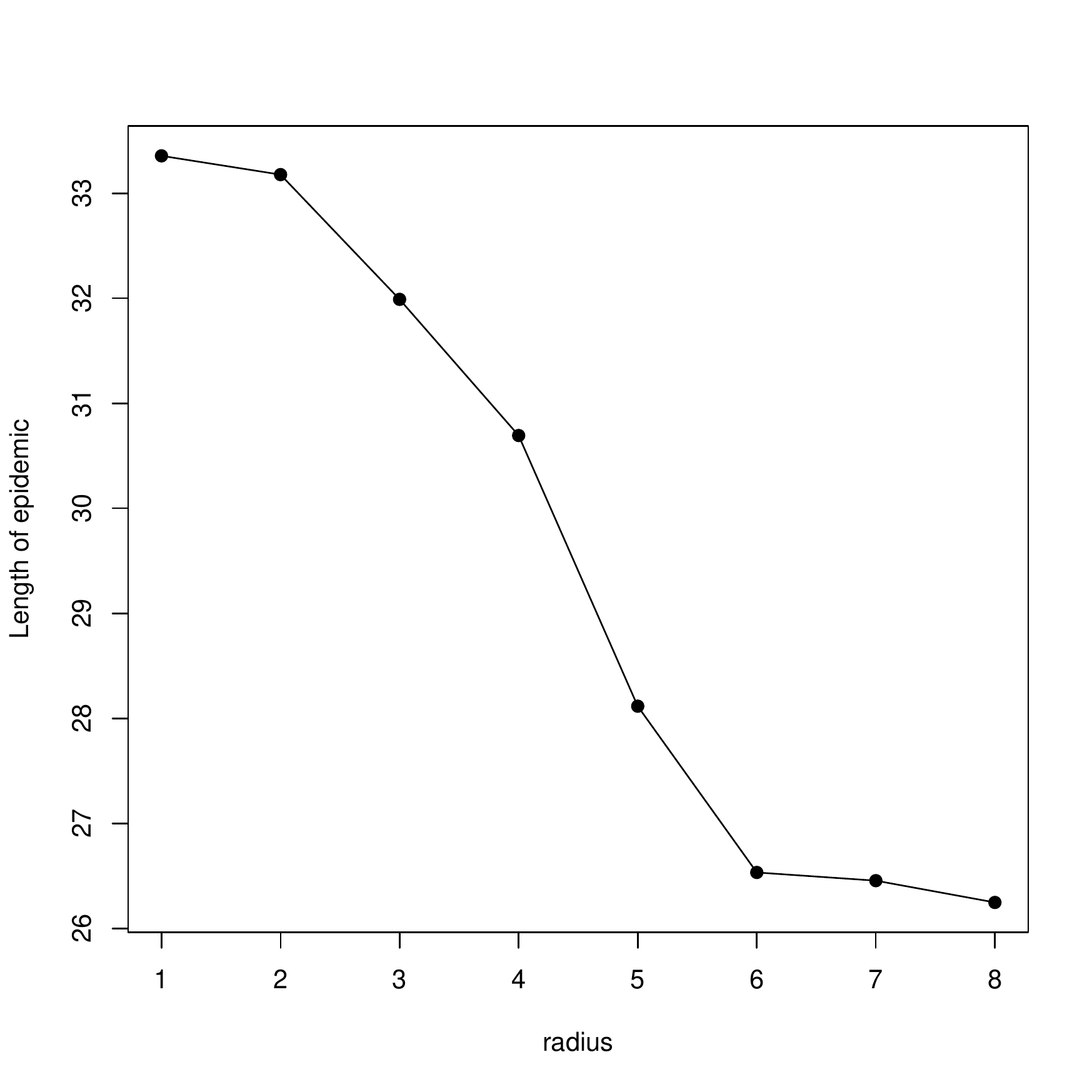}
\caption{The average length of epidemics at each radius.}
\label{length-epi}
\end{center}
\end{figure}

Figure \ref{infec-culled} shows the average number of infected and culled individuals at each radius. We can see that the number of infected individuals tends to decrease dramatically as the radius of the ring increases, levelling off once we have to get around $r = 5$ units. However, the number of culled individuals also increases quite dramatically with increasing the radius of the ring, also levelling off around $r = 7$ units. We can also see from Figure \ref{length-epi} increasing the radius $r$ tends to decrease the length of the epidemic.

Of course, the \code{control.strategy} function can be easily modified to impose other control strategies. For example, instead of culling within a time step as in the case here, we could allow for (stochastic) delays between infection and culling for surrounding individuals, or allow for only a probability of failure regarding each cull or vaccination event.

\newpage

\section[D.]{Comparing the computation times of running different models}\label{appendix.4}

Here, we compare the effect of population size and the number of infected individuals on the computation time for running the \code{epictmcmc} function. We considered five population sizes (50, 250, 450, 650 and 850 individuals), and generated three different epidemics using $\mathcal{SIR}$ distance-based continuous time ILMs, via the \code{datagen} function, resulting in different numbers of infected individuals. These epidemics are categorized into three levels as: small, medium, large defined as epidemics in which the number of infected individuals are less than 25\%, between 25\% and 50\%, or greater than 50\% of the population, respectively. Then, we run the \code{epictmcmc} three times assuming \code{datatype} = \code{"known epidemic"}, \code{"known removal"} with single chain, and \code{"known removal"} with three chains, updating the infection times in blocks of size five. 

\begin{figure}[!h]
\begin{center}
\includegraphics[width=\textwidth,height=10cm]{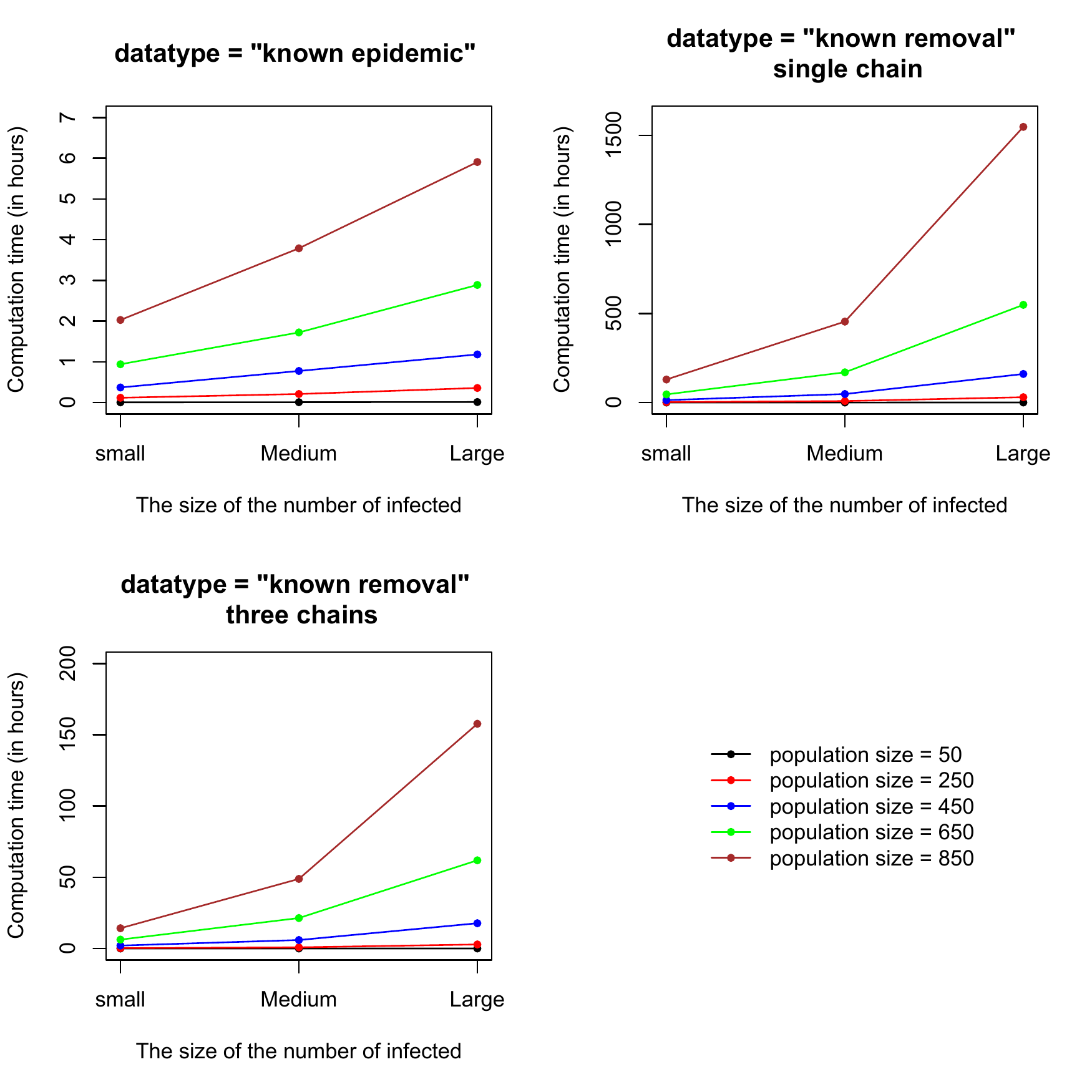}
\caption{Approximated computation times of running the \code{epictmcmc} function for fitting different epidemic data sets, with different population sizes and number of infected individuals, using $\mathcal{SIR}$ distance-based continuous time ILMs under three scenarios: fully observed epidemics, partially observed epidemics with a single MCMC chain, and three MCMC chains.}
\label{comp-time-different}
\end{center}
\end{figure}

Figure~\ref{comp-time-different} shows the computation times in hours for running the \code{epictmcmc} function on the above epidemics to obtain 150,000 MCMC samples. The computation times were approximated on the basis of running ten iterations, as our concern here is just to see to estimate the effect of population size and number of infected individuals upon computation time. We observed strong correlation between the population sizes and number of infected individuals in all of the considered analysis scenarios. 

We see that under the fully observed epidemic assumption (\code{datatype}= \code{"known epidemic"}), the function \code{epictmcmc} can be performed in reasonable time for all scenarios. However, computation time becomes an issue for partially observed epidemics (\code{datatype}= \code{"known removal"}) when updating the infection times in turn in a single chain. Larger epidemics with larger population sizes were estimated to take more than four weeks to obtain 150,000 MCMC samples. Computation time is greatly reduced by running \code{epictmcmc} over multiple chains and updating infection times in blocks of size five. For example, with an epidemic in a population of size was 850 individuals, and almost all of individuals infected, the computation time was reduced from approximately 1548 hours ($\approx$65 days) to 157 hours ($\approx$7 days).

\end{document}